\documentclass[11pt]{article}

\usepackage{amsmath}
\usepackage{amssymb}
\usepackage{amsthm}
\usepackage{fullpage}
\usepackage[draft=false,hypertexnames=false,bookmarksopenlevel=1,bookmarksopen,bookmarksnumbered,colorlinks,plainpages=false,linktocpage,linkcolor=black,citecolor=black]{hyperref}
\usepackage{enumerate}
\usepackage{graphicx}
\usepackage{xcolor}
\usepackage{mathtools}
\usepackage{dsfont}
\usepackage[bottom]{footmisc}
\usepackage{bbm}
\usepackage{comment}
\usepackage{tikz}
\usetikzlibrary{decorations.pathreplacing}

\newcommand{\R}{\mathbb{R}}

\newcommand{\Z}{\mathbb{Z}}

\newcommand{\C}{\mathbb{C}}
\newcommand{\F}{\mathbb{F}}
\newcommand{\eps}{\varepsilon}
\newcommand{\coeff}{\mathrm{COEF}}
\newcommand{\paren}[1]{(#1)}  

\newcommand{\chevron}[1]{\langle #1 \rangle}
\newcommand{\abs}[1]{\left\lvert #1 \right\rvert}
\newcommand{\floor}[1]{\left\lfloor #1 \right\rfloor}
\newcommand{\ceil}[1]{\left\lceil #1 \right\rceil}

\theoremstyle{plain}
\newtheorem{theorem}{Theorem}[section]
\newtheorem{thm}[theorem]{Theorem}
\newtheorem*{thmrest}{Theorem \ref{thm:the-main} \normalfont(restated)}

\newtheorem{la}[theorem]{Lemma}

\newtheorem{cor}[theorem]{Corollary}
\newtheorem{proposition}[theorem]{Proposition}
\newtheorem{prop}[theorem]{Proposition}

\theoremstyle{definition}
\newtheorem{definition}[theorem]{Definition}
\newtheorem{df}[theorem]{Definition}
\newtheorem{notation}[theorem]{Notation}

\newtheorem{ex}[theorem]{Example}

\newtheorem{rmk}[theorem]{Remark}

\newcommand{\wt}{\widetilde}
\newcommand{\wh}{\widehat}

\newcommand{\Path}{\mr{Path}}
\newcommand{\IMM}{\mr{IMM}}

\newcommand{\transpose}{\mathsf{T}}
\usepackage{scalerel}
\newcommand{\kron}{\mathrel{\scalerel*{\boxtimes}{\otimes}}}
\newcommand{\bigkron}{\mathop{\scalerel*{\boxtimes}{\big|}}}

\def\[#1\]{\begin{align*}#1\end{align*}}

\newcommand{\mc}{\mathcal}
\newcommand{\mr}{\mathrm}

\newcommand{\tu}{\textup}

\newcommand{\defeq}{\vcentcolon=}

\newcommand{\tri}{\vartriangle}
\newcommand{\BAR}[1]{\overline{#1}}

\newcommand{\sq}[1]{\ensuremath{\langle#1\rangle}}
\newcommand{\relrk}{\mr{relrk}}
\newcommand{\rank}{\mr{rank}}
\newcommand{\PTrank}{\mr{PT\textup{-}rank}}

\newcommand{\SOS}{\mr{SoS}}

\newcommand{\Lsm}{\mc{L}_{\mathsf{sm}}}
\newcommand{\Lnc}{\mc{L}_{\mathsf{nc,sm}}}

\newcommand{\VNCone}{\mathsf{VNC^1}}
\newcommand{\VBP}{\mathsf{VBP}}
\newcommand{\VP}{\mathsf{VP}}
\newcommand{\VNP}{\mathsf{VNP}}
\renewcommand{\P}{\mathsf{P}}
\newcommand{\NP}{\mathsf{NP}}

\newcommand{\VBPord}{\mathsf{VBP_{ord}}}

\newcommand{\VPnc}{\mathsf{VP}_{\mathsf{nc}}}
\newcommand{\VNPnc}{\mathsf{VNP}_{\mathsf{nc}}}

\newcommand{\Mat}{\mr{Mat}}
\newcommand{\ShiftT}[1]{\mathrm{ShiftedTensor}(#1)}
\newcommand{\PaddedT}[1]{\mathrm{PaddedTensor}(#1)}

\title{Multiquadratic Sum-of-Squares Lower Bounds Imply VNC$^1$ $\neq$ VNP}

\author{Benjamin Rossman \\ Duke University \and Davidson Zhu \\ Duke University}

\date{\today}

\begin{document}
\maketitle

\begin{abstract}
The \emph{sum-of-squares (SoS) complexity} of a $d$-multiquadratic polynomial $f$ (quadratic in each of $d$ blocks of $n$ variables) is the minimum $s$ such that $f = \sum_{i=1}^s g_i^2$ with each $g_i$ $d$-multilinear.
In the case $d=2$,
Hrube\v{s}, Wigderson and Yehudayoff~\cite{hrubevs2011non} showed that an $n^{1+\Omega(1)}$ lower bound on the SoS complexity of explicit biquadratic polynomials implies an exponential lower bound for non-commutative arithmetic circuits.
In this paper, we establish an analogous connection between general \emph{multiquadratic sum-of-squares} and \emph{commutative arithmetic formulas}. 
Specifically, we show that an $n^{d-o(\log d)}$ lower bound on the SoS complexity of explicit $d$-multiquadratic polynomials, for any $d = d(n)$ with $\omega(1) \le d(n) \le O(\frac{\log n}{\log\log n})$, would separate the algebraic complexity classes $\VNCone$ and $\VNP$.
\end{abstract}

\newpage

\tableofcontents{}

\newpage

\section{Introduction}\label{sec:intro}

Computational complexity theory aims to categorize computational problems by the amount of time and resources they need. The algebraic complexity classes proposed by Leslie Valiant \cite{valiant1979completeness}, on the other hand, were created for understanding the tractability of polynomials, as opposed to Boolean functions. Arithmetic circuits, branching programs and formulas are defined as analogs of their Boolean counterparts. 

Algebraic complexity classes $\mathsf{VF}\ (=\VNCone)
\subseteq \VBP \subseteq \VP\ (=\mathsf{VNC^2}) \subseteq \VNP$ consist of sequences of $\mr{poly}(n)$-degree $n$-variate polynomials that are computable by $\mr{poly}(n)$-size arithmetic formulas, branching programs, circuits and projections of circuits, respectively.  As with the analogous Boolean classes ($\mathsf{NC^1} \subseteq \mathsf{NL} \subseteq \mathsf{P} \subseteq \mathsf{NP}$), these four algebraic classes are believed to be distinct.
However, it remains a major open problem even to separate $\VNCone$ from $\VNP$, despite significant progress on lower bounds for formulas, branching programs and circuits in various restricted settings: low-depth \cite{amireddy2023low,limaye2022partial,limaye2022superpolynomial}, multilinear \cite{dvir2012separating,nisan1996lower,raz2004multilinear}, set-multilinear \cite{arvind2016some,kush2023near,raz2008balancing}, and non-commutative \cite{chatterjee2021separating,nisan1991lower,tavenas2022set} (to mention just a few representative works).

\subsection{Biquadratic sum-of-squares and non-commutative circuits}

Let $f \in \F[X_1,\dots,X_n,Y_1,\dots,Y_n]$ be a biquadratic polynomial over a field $\F$ with monomials of the form $X_{i_1} X_{i_2} Y_{j_1} Y_{j_2}$. 
The {\em (biquadratic) sum-of-squares complexity} of $f$, denoted $\SOS(f)$, is the minimum number $s$ of bilinear polynomials $g_1,\dots,g_s$ such that $f = g_1^2 + \dots + g_s^2$.
When $\F$ is algebraically closed, we have $\SOS(f) \le n^2$ for all $f$, while a dimension-counting argument shows that $\SOS(f) = \Omega(n^2)$ for almost all $f$.

Sum-of-squares complexity has been intensively studied for the separable biquadratic polynomial
\[
  q_n \defeq (X_1^2 + \dots + X_n^2) \cdot (Y_1^2 + \dots + Y_n^2).
\]
Over $\R$ and $\C$, trivial bound are given by
\[
  n \le \SOS_\C(q_n) \le \SOS_\R(q_n) \le n^2.
\]
A classical theorem of Hurwitz from 1898 shows that $\SOS_{\R}(q_n) = n$ iff $n \in \{1,2,4,8\}$ \cite{hurwitz1922komposition}.
An upper bound $\SOS_{\R}(q_n) = O(\frac{n^2}{\log n})$ is given by the classical Radon-Hurwitz construction (see \cite{rajwade1993squares,shapiro2011compositions}).
Asymptotically, this remains the strongest known upper bound over $\R$.
It was also the strongest known bound over $\C$, until a recent breakthrough of Hrube\v{s}~\cite{hrubes2024subquadratic}:

\begin{thm}[\cite{hrubes2024subquadratic}]\label{thm:Hrubes}
There exists a sum-of-squares decomposition of $q_n$ over $\C$ with just $O(n^{1.62})$ terms, that is, $\SOS_\C(q_n) = O(n^{1.62})$. 
\end{thm}

As for lower bounds, Adams and Atiyah \cite{adams1966k} showed 
$\SOS_\C(q_n) \ge 2^{\lceil\log_2 n\rceil}$; 
in particular, 
$\SOS_\C(q_n) \ge 2n-2$ infinitely often. Slightly better bounds have been obtained for specific values of $n$ (and for general fields of characteristic $\neq 2$) 
using techniques from algebraic K-theory 
\cite{dugger2007hopf}. However, no lower bound better than $\SOS_\C(q_n) = \Omega(n)$ is currently known.

A striking result of Hrube\v{s}, Wigderson and Yehudayoff~\cite{hrubevs2011non} explains the difficulty of improving this lower bound.
They show that proving a sufficiently super-linear lower bound on $\SOS_\C(q_n)$ would have a dramatic consequence in non-commutative algebraic circuit complexity:

\begin{thm}[\cite{hrubevs2011non}]\label{thm:HWY}
Let $f = (f_n)$ be any explicit sequence\footnote{The sequence $f = (f_n)$ is {\em explicit} if the coefficient in $f_n$ of each monomial $X_{i_1}X_{i_2}Y_{j_1}Y_{j_2}$ is computable in $\mr{polylog}(n)$ time given $n$ and indices $i_1,i_2,j_1,j_2 \in [n]$.} of biquadratic polynomials over $\C$.
If $\SOS_\C(f) = \Omega(n^{1+\eps})$ for any $\eps > 0$, then the permanent $\mathrm{PERM}_n$ requires exponential size non-commutative circuits; 
in particular, this implies a separation of complexity classes $\VPnc$ and $\VNPnc$ (the non-commutative algebraic analogues of $\P$ and $\NP$).
Moreover, a lower bound $\SOS_\C(f) = \Omega(n \log^5 n)$ suffices for the separation $\VPnc \neq \VNPnc$.
\end{thm}

Notice that Hrube\v{s}'s upper bound $\SOS_\C(q_n) = O(n^{1.62})$ does not rule out the possibility of separating $\VPnc$ and $\VNPnc$ via an $\Omega(n^{1+\eps})$ lower bound.
(We remark that Theorems \ref{thm:Hrubes} and \ref{thm:HWY} 
hold more generally over any algebraically closed field, but are not known to extend to $\R$; see the discussion in \S\ref{sec:related}.)

\subsection{Multiquadratic sum-of-squares and commutative set-multilinear formulas}

The main theorem of this paper shows that strong enough lower bounds for {\em multiquadratic sum-of-squares} imply super-polynomial lower bounds for {\em commutative arithmetic formulas}.

Let $f$ be a $d$-multiquadratic polynomial over an algebraically closed field, that is, a linear combination of monomials $X^{(1)}_{i_1}X^{(1)}_{j_1} \cdots X^{(d)}_{i_d}X^{(d)}_{j_d}$ where $i_1, j_1, \dots, i_d, j_d \in [n]$. 
As in the biquadratic setting ($d=2$), the {\em (multiquadratic) sum-of-squares complexity} of $f$, denoted $\SOS(f)$, is defined as the minimum number $s$ of $d$-multilinear polynomials $g_1,\dots,g_s$ such that $f = g_1^2 + \dots + g_s^2$. 
Similar to the biquadratic setting, we have $\SOS(f) \le O(n^d)$ for all $f$, and $\SOS(f) = \Omega((n/2)^d)$ for almost all $f$ (Prop.~\ref{prop:fully-symmetric}). 

Our main result---stated informally below and formally in the next subsection---shows that a nearly optimal lower bound on the SoS complexity of explicit $d$-multiquadratic polynomials would separate the algebraic complexity classes $\VNCone$ and $\VNP$.

\begin{thm}[informal]\label{thm:informal}
Let $f = (f_n)$ be any explicit sequence of $d$-multiquadratic polynomials over an algebraically closed or finite field of characteristic $\neq 2$. If $\SOS(f) = n^{d-o(\log d)}$, then this yields an $n^{\Omega(\log d)}$ lower bound on the set-multilinear formula size of associated $d+1$-multilinear polynomials $\wt f = (\wt f_n)$.
If in addition $d = d(n)$ satisfies $\omega(1) \le d(n) \le O(\frac{\log n}{\log\log n})$, we get a separation $\VNCone \neq \VNP$ (via a lemma of Raz~\cite{raz2013tensor}).
\end{thm}

While Theorem~\ref{thm:informal} is analogous to the result of Hrube\v{s}, Wigderson and Yehudayoff~\cite{hrubevs2011non}, it is closer in nature to a well-known theorem of Raz~\cite{raz2013tensor} linking {\em tensor rank} to set-multilinear formula lower bounds (see the discussion in \S\ref{sec:related}).

\subsection{Partial transpose rank of $n^d \times n^d$ matrices}

\begin{definition}
    Let $\mc{X}=\{X_{i_1}^{(1)}\}_{i_1=1}^n\cup \{X_{i_2}^{(2)}\}_{i_2=1}^n\cup\cdots\cup\{X_{i_d}^{(d)}\}_{i_d=1}^n$ be $d$ disjoint blocks of $n$ variables. A polynomial $P(\mc{X})$ is $d$-multiquadratic if each monomial of $P(\mc{X})$ has total degree exactly $2$ in each block. 
\end{definition}

Every $n^d \times n^d$ matrix $M$ gives rise to a $d$-multiquadratic polynomial defined by
\[
  Q_M \defeq \sum_{\vec i,\vec j \in [n]^d} M_{\vec i,\vec j} \cdot X^{(1)}_{i_1}X^{(1)}_{j_1} \cdots X^{(d)}_{i_d}X^{(d)}_{j_d}.
\]
Conversely, for fields of characteristic $\neq 2$, we get a bijective correspondence between $d$-multiquadratic polynomials (with $d$ blocks of $n$ variables) and $n^d \times n^d$ matrices satisfying 
$M = M^{\top_1} = \dots = M^{\top_d}$, where ${\top_k}$ is the $k$-th partial transpose operation which swaps row-index $i_k$ and column-index $j_k$.  We refer to such matrices as {\em fully symmetric}.

The actual main theorem of this paper concerns a generalization of SoS complexity called {\em partial transpose rank}---or {\em PT-rank} for short---which is a complexity measure on general $n^d \times n^d$ matrices (not just fully symmetric ones).
We say that an $n^d \times n^d$ matrix $P$ is {\em PT-basic} if 
$P^{\top_\kappa}$ has rank $1$ for any $\kappa \subseteq [d]$, where ${}^{\top_\kappa}$ is the composition of (commuting) operations ${\top_k}$ for $k \in \kappa$.
We then define the {\em PT-rank} of any $n^d \times n^d$ matrix $M$ as the minimum number $r$ of PT-basic matrices $P_1,\dots,P_r$ such that $M = P_1+\dots+P_r$.
This complexity measure is closely related to SoS complexity:
for fully symmetric matrices $M$ (satisfying $M = M^{\top_k}$ for all $k \in [d]$), we show that $\PTrank(M)$ and $\SOS(Q_M)$ are equivalent to an $O(2^d)$ factor over any algebraically closed \underline{or} finite field of characteristic $\neq 2$ (Prop.~\ref{prop:pt-rank-bil}).

The actual main theorem of this paper relates the PT-rank of an arbitrary $n^d \times n^d$ matrix $M$ with the set-multilinear formula size of an associated multilinear polynomial of degree $d+1$ in $dn^2$ ($= dn+(d-1)n^2+dn$) variables:
\[
  \wt Q_M \defeq 
  \sum_{\vec i,\vec j \in [n]^d} M_{\vec i,\vec j} \cdot  
  \wt X^{(0)}_{i_1} \wt X^{(1)}_{j_1,i_2} \cdots \wt X^{(d-1)}_{j_{d-1},i_d} \wt X^{(d)}_{j_d}.
\]

\begin{thm}[Main theorem]\label{thm:the-main}
For every $n^d \times n^d$ matrix $M$ (over any field), the set-multilinear formula size of $\wt Q_M$ is at least $\PTrank(M)/n^{d - \log d + 1}$.
\end{thm}

Via a standard set-multilinearization lemma of Raz~\cite{raz2013tensor}, we get the following corollary:

\begin{cor}\label{cor:the-main}
Suppose that $\PTrank(M) = n^{d - o(\log d)}$ where $M$ is any explicit sequence of $n^d \times n^d$ matrices with $\omega(1) \le d(n) \le O(\frac{\log n}{\log\log n})$.
Then $\wt Q_M$ has arithmetic formula size at least $n^{(1-o(1))\log d}$ and, hence, $\VNCone \neq \VNP$.
\end{cor}

(Theorem \ref{thm:informal} on SoS complexity follows directly from Theorem \ref{thm:the-main} and Corollary \ref{cor:the-main}. For a $d$-multiquadratic polynomial $f = Q_M$, the multilinear polynomial $\wt f$ of Theorem \ref{thm:informal} is given by $\wt Q_M$.)

Since almost all $M$ have PT-rank $\ge (\frac12-o(1))n^d$ (Prop.\ \ref{prop:almost-all}), finding an explicit $M$ with $\PTrank(M) = n^{d - o(\log d)}$ is a classic problem of ``finding hay in a haystack''.
By finding a very explicit $M$ (with $\wt Q_M$ in $\VP$ rather than $\VNP$), Corollary \ref{cor:the-main} could conceivably be used to show $\VNCone \neq \VP$. 
On the flipside, an additional result of this paper (Theorem~\ref{thm:limitation-theorem}) rules out lower bounds, over fields of characteristic $\neq 2$, whenever $\wt Q_M$ belongs to the class $\VBPord$ of set-multilinear polynomials computable by a polynomial-size ordered branching program. 
This negative result leverages the recent upper bound $\SOS_{\C}(q_n) = O(n^{1.62})$ of Hrube\v{s}~\cite{hrubes2024subquadratic} (Theorem \ref{thm:Hrubes}).
We also show that $\PTrank(M)$ is at most $n^{O(d^{0.7})}$ when $M$ is a Kronecker product of $d$ $n \times n$ matrices over any fields of characteristic $\ne 2$ (Theorem \ref{thm:kron}); this observation rules out certain explicit $n^d \times n^d$ matrices as candidates for having PT-rank $n^{d-o(\log d)}$.

So far as we know, over $\mr{GF}(2)$, it is possible that the $n^d \times n^d$ {\em identity matrix} $I_{n^d}$ has PT-rank $n^{d-o(\log d)}$ or even exactly $n^d$.  If so, by Theorem \ref{thm:the-main} this implies an $n^{(1-o(1))\log d}$ lower bound on the formula size of {\em iterated matrix multiplication} $\mr{IMM}_{n,d+1}$ ($= \wt Q_{I_{n^d}}$), which would separate classes $\VNCone$ and $\VBP$ over $\mr{GF}(2)$.


\subsection{Related work}\label{sec:related}

Following the influential paper of Hrube\v{s}, Wigderson and Yehudayoff~\cite{hrubevs2011non}, subsequent work by various authors \cite{arvind2016some,carmosino2018hardness,dutta2024weighted} has found additional reductions between various sum-of-squares complexity measures and algebraic circuit lower bounds.  Most recently, Dutta, Saxena and Thierauf \cite{dutta2024weighted} showed that an $d^{0.5 + \Omega(1)}$ lower bound on the support-sum of any explicit degree-$d$ univariate polynomial would separate $\VP$ and $\VNP$.
However, none of the prior work has explored the $d$-multiquadratic setting, nor implications specific to arithmetic formulas. 

More closely related to our main theorem is the well-known theorem of Raz~\cite{raz2013tensor} that an $n^{d-o(d)}$ lower bound on the {\em tensor rank} of any explicit $d$-multilinear polynomial $g$ (i.e.\ order-$d$ tensor $[n]^d \to \F$) implies an $n^{\Omega(\log d)}$ lower bound on the set-multilinear formula size of the same polynomial $g$.
The further implication $\VNCone \neq \VNP$ when $\omega(1) \le d(n) \le O(\frac{\log n}{\log\log n})$ follows from an additional lemma in \cite{raz2013tensor}, which we also invoke above.

We remark that a lower bound $\SOS(f) = n^{d-o(\log n)}$ for an explicit $d$-multiquadratic polynomial $f$ (i.e.\ fully symmetric $n^d \times n^d$ matrix) is conceivable without having to break the longstanding $O(n^{\lfloor d/2 \rfloor})$ tensor rank barrier.
Indeed, Theorem \ref{thm:informal} appears to evade the known barriers for rank-based lower bound methods identified by Efremenko {\em et al} \cite{efremenko2018barriers} and Garg {\em et al} \cite{garg2019more}.

\subsection{Outline of the paper}

The rest of the paper is organized as follows:
\S\ref{sec:prelims} presents key definitions pertaining to tensors, matrices, and arithmetic formulas.
\S\ref{sec:PT-rank} formally introduces {\em partial transpose rank} and explores its properties.
\S\ref{sec:warm-up} proves a weaker (but easier and illustrative) version of our main theorem---with respect to {\em non-commutative} set-multilinear formulas.
\S\ref{sec:proof} proves the main theorem---with respect to {\em commutative} set-multilinear formulas.
\S\ref{sec:VBP} discusses upper bounds for PT-rank and candidate explicit hard matrices.

\section{Preliminaries}\label{sec:prelims}

Throughout this paper, $n,d$ are arbitrary positive integers, where we regard $n$ as a growing parameter and $d = d(n)$ as a function of $n$.  

We write $[n]$ for the set $\{1,\dots,n\}$ and $\log n$ for the base-2 logarithm of $n$.
We write $\sq{\cdot,\cdot} : [n]^2 \to [n^2]$ for the pairing bijection defined by $\sq{i,j} \defeq (i-1)n + j$.

\subsection{The underlying field}

Throughout this paper, $\F$ is an arbitrary field. We will explicitly state whenever an assumption on $\F$ is required. In particular, our main theorem (Theorem \ref{thm:the-main}) and other results concerning PT-rank work for completely general fields (including characteristic $2$). 

The only results that depend on $\F$ concern the relationship between PT-rank and SoS complexity (Lemma \ref{prop:pt-rank-bil}) and the consequence of our main theorem stated in the terms of SoS complexity (Theorem \ref{thm:informal}). Even here, our results apply to a large class of fields of characteristic $\neq 2$, including all algebraically closed fields and finite fields.

\subsection{Matrices and tensors}

In the Introduction, our main theorem was stated first in terms of SoS complexity of {\em $d$-multiquadratic polynomials}, then more generally in terms of PT-rank of {\em $n^d \times n^d$ matrices}. These rank measures were linked to the algebraic complexity of a third kind of object: {\em $d+1$-multilinear polynomials}.
It is convenient to regard these objects as different flattenings of a tensor $[n]^{2d} \to \F$ to tensors of format $[n^2]^d$ or $[n^d]^2$ or $[n] \times [n^2]^{d-1} \times [n]$, respectively.

\begin{df}[Tensors]
An {\em order $d$ tensor} is a function $A : [n_1] \times \dots \times [n_d] \to \F$ where $n_1,\dots,n_d$ are nonnegative integers.\footnote{All definitions and claims in this paper can be rephrased in a basis-free manner, that is, regarding $A$ as a multilinear function $\F^{n_1} \times \dots \times \F_{n_d} \to \F$ without reference to any specific bases for vector spaces $\F^{n_i}$.}
We shall mainly (but not exclusively) consider tensors $[n]^d \to \F$ where $n_1=\dots=n_d=n$.
More generally, we consider tensors $A : [n]^D \to \F$ with coordinates indexed by a nonempty finite set $D$.\footnote{Specifically, in \S\ref{sec:proof} we consider tensors where $D$ is the directed edge set of an arbitrary subgraph of $\Path_d$.}

For tensors $A : [n]^d \to \F$ and $B : [n]^e \to \F$, their tensor product is denoted by $A \otimes B : [n]^{d+e} \to \F$.
For tensors $A : [n]^D \to \F$ and $B : [n]^E \to \F$ with $D$ and $E$ disjoint, their tensor product is denoted by $A \otimes B : [n]^{D \cup E} \to \F$. 

The {\em tensor rank} of a tensor $A : [n_1] \times \dots \times [n_d] \to \F$, denoted $\textrm{tensor-rank}(A)$, is the minimum number $r$ such that $A$ admits a decomposition $A = \sum_{j=1}^r (A_{1,j} \otimes \dots \otimes A_{d,j})$ for some collection of order 1 tensors $A_{i,j} : [n_i] \to \F$.\footnote{
We define {\em tensor rank} solely for the sake of comparing our main result (Theorem \ref{thm:the-main}) with Raz's theorem shows that an $n^{d-o(d)}$ lower bound on tensor rank implies $\VNCone \neq \VP$ \cite{raz2013tensor}, as discussed in \S\ref{sec:related}. Elsewhere in this paper, we only every consider the matrix rank of flattened tensors. 
}
\end{df}

\begin{df}[Matrices]
A matrix is an order $2$ tensor $M : [m_1] \times [m_2] \to \F$, also written as $M \in \F^{m_1 \times m_2}$.
The {\em rank} of $M$ is denoted by $\rank(M)$; note that this is a special case tensor rank. The {\em transpose} of $M$ is denoted by $M^\transpose \in \F^{m_2 \times m_1}$. 
(We use a bolder font $\transpose$ to distinguish the full transpose from partial transpose operations $\top_\kappa$ on $n^d \times n^d$ matrices introduced in Def.~\ref{df:top}.)

For matrices $M : [m_1] \times [m_2] \to \F$ and $N : [n_1] \times [n_2] \to \F$, their Kronecker product is denoted by $M \kron N : [m_1n_1] \times [m_2n_2] \to \F$ (not to be confused with the tensor product $M \otimes N : [m_1] \times [m_2] \times [n_1] \times [n_2] \to \F$.)

The $n \times n$ identity matrix is denoted by $I_n$.  We write $I_n^{\otimes d}$ for the order $2d$ tensor $I_n \otimes \dots \otimes I_n$ ($d$ times) (not to be confused with the $n^d \times n^d$ matrix $I_{n^d} = I_n \kron \dots \kron I_n$, which is a tensor of order $2$).

When $M$ is an $n^d \times n^d$ matrix, we index its rows and columns by $d$-tuples in $[n]^d$.
That is, we express the entries of $M$ as $M_{(i_1,\dots,i_d),(j_1,\dots,j_d)}$.
\end{df}

\begin{df}[Matrix flattening of a tensor]
For a tensor $A : [n]^D \to \F$ and partition $I \sqcup J = D$, we denote $\Mat_{I,J}(A)$ for the flattened matrix
\[
  \Mat_{I,J}(A) : [n^{|I|}] \times [n^{|J|}] \to \F.
\]
\end{df}

\subsection{Set-multilinear formula size}\label{sec:p_A}

There is a one-to-one correspondence between order $d$ tensors $A : [n_1] \times \dots \times [n_d] \to \F$ and set-multilinear polynomials $P_A \in \F_{\mathsf{sm}}[\mc X]$ in variables $\mc X = \{X^{(i)}_a : i \in [d],\, a \in [n_i]\}$.  Namely, $A$ describes the coefficients of $P_A$:
\[
  P_A(\mc X) = \sum_{(a_1,\dots,a_d) \in [n_1] \times \dots \times [n_d]} A_{a_1,\dots,a_d}\ X^{(1)}_{a_1} \cdots X^{(d)}_{a_d}.
\]
For any nonempty finite set $D$, we have a similar correspondence between tensors $A : [n]^D \to \F$ and set-multilinear polynomial $P_A \in \F_{\mathsf{sm}}[\mc X]$ in variables $\mc X = \{X^{(i)}_a : i \in D,\, a \in [n]\}$.  

\begin{df}[Set-multilinear formula size]
The {\em set-multilinear formula size} of a tensor $A : [n]^D \to \F$, denoted $\Lsm(A)$, is the minimum number of leaves in a syntactically set-multilinear arithmetic formula that computes the associated polynomial $P_A$.
This may also be defined directly by induction:
\begin{itemize}
  \item
    If $|D| = 1$, then $\Lsm(A)$ is $1$ if $A$ is non-zero anywhere and $0$ otherwise.
  \item
    If $|D| \ge 2$, then $\Lsm(A)$ is the minimum value of $\sum_i\ (\Lsm(B_i) + \Lsm(C_i))$
    over indexed families $\{(E_i,F_i,B_i,C_i)\}_i$ (where $i$ runs over an arbitrary, unnamed index set) such that $E_i,F_i$ are disjoint nonempty sets satisfying $E_i \cup F_i = D$ and $B_i : [n]^{E_i} \to \F$ and $C_i : [n]^{F_i} \to \F$ are tensors satisfying $A = \sum_i B_i \otimes C_i$.  
\end{itemize}
\end{df}

For further background on algebraic complexity measures, see  \cite{burgisser2013algebraic,saptharishi2015survey,shpilka2010arithmetic}.

\subsection{$n^d \times n^d$ matrices and their shifted tensors}\label{sec:matrices-shifted-tensors}

Recall that our main theorem, Theorem \ref{thm:the-main}, relates $\PTrank(M)$ to the set-multilinear formula size of a related polynomial. This polynomial is denoted $\wt Q_M$ in the statement of Theorem \ref{thm:the-main}. However, it is convenient to speak of the associated tensor, which we shall refer to as the ``shifted tensor'' of $M$.

\begin{df}[The shifted tensor of an $n^d \times n^d$ matrix $M$]\label{df:shift}
For any $n^d \times n^d$ matrix $M$, we defined the reshaped tensor $\ShiftT{M} : [n^2]^{d+1} \to \F$ as follows. 
Recall that $\sq{\cdot,\cdot}$ is a bijection $[n]^2 \to [n^2]$.
For all $p,i_1,j_1,\dots,i_d,j_d,q \in [n]$, we have
\begin{align*}
  \ShiftT{M}(\sq{p,i_1},\sq{j_1,i_2},\dots,\sq{j_{d-1},i_d},\sq{j_d,q}) 
  &\defeq
  \mathds 1\{p=q=1\} \cdot
  M_{(i_1,\dots,i_d),(j_1,\dots,j_d)}.
\end{align*}
Note that the set-multilinear polynomial $P_{\ShiftT{M}}$ is precisely $\wt Q_M$ of Theorem \ref{thm:the-main}. 
\end{df}

\begin{df}[Iterated Matrix Multiplication]
An important example of a ``shifted tensor'' is the {\em Iterated Matrix Multiplication} tensor $\IMM_{n,d} : [n^2]^d \to \F$ defined by
\[
  \IMM_{n,d}(\sq{a_1,b_1},\dots,\sq{a_d,b_d})
  \defeq
  \mathds 1[\,(b_1,\dots,b_{d-1})=(a_2,\dots,a_d)\,].
\]
The familiar corresponding set-multilinear polynomial describes the sum of entries in the product of $d$ symbolic $n \times n$ matrices:
\[
  P_{\IMM_{n,d}}(\mc X)
  =
  \sum_{a_0,\dots,a_d \in [n]}
  X^{(1)}_{\sq{a_0,a_1}} X^{(2)}_{\sq{a_1,a_2}} \cdots X^{(d)}_{\sq{a_{d-1},a_d}}. 
\]
\end{df}

Note that $\IMM_{n,d} = \ShiftT{I_{n^{d-1}}}$, that is, $\IMM_{n,d}$ is the shifted tensor of the $n^{d-1} \times n^{d-1}$ identity matrix.

\section{Partial transpose rank}\label{sec:PT-rank}

The partial transpose is a key concept in quantum information theory, particularly in studying entanglement and separability of quantum states 
\cite{horodecki1996necessary,peres1996separability}.
In this section, we introduce the {\em partial transpose rank} of an $n^d \times n^d$ matrix.  
To the best of our knowledge, this notion has not been explicitly studied before. As we show, PT rank generalizes SoS complexity (for fully symmetric matrices over algebraically closed or finite fields of characteristic $\neq 2$).

We begin by defining the {\em partial transpose} operations on $n^d \times n^d$ matrices.

\begin{df}[Partial transpose]\label{df:top}
Let $M$ be an $n^d \times n^d$ matrix.
\begin{itemize}
\item
For $k \in [d]$, the {\em $k$-th partial transpose} of $M$, denoted $M^{\top_k}$, is the $n^d \times n^d$ matrix defined by
\[
  M^{\top_k}_{(i_1,\dots,i_d),(j_1,\dots,j_d)}
  \defeq
  M^{\vphantom{\top_k}}_{(i_1,\dots,i_{k-1},j_k,i_{k+1},\dots,i_d),(j_1,\dots,j_{k-1},i_k,j_{k+1},\dots,j_d)}.
\]
\item
For $\kappa \subseteq [d]$, the {\em $\kappa$-partial transpose} of $M$ is the matrix $M^{\top_\kappa} \defeq M^{\top_{k_1}\dots \top_{k_c}}$ where $k_1,\dots,k_c$ are the distinct elements of $\kappa$ (in any order).
\end{itemize}
\end{df}

Note that $M^{\top_\emptyset} = M$ and $M^{\top_{[d]}} = M^\transpose$ (the usual transpose of $M$). Also note that 
\[
M^{\top_{\kappa_1}\top_{\kappa_2}} =
M^{\top_{\kappa_2}\top_{\kappa_1}} = 
M^{\top_{\kappa_1 \tri \kappa_2}}
\]
where $\kappa_1 \tri \kappa_2 = (\kappa_1 \setminus \kappa_2) \cup (\kappa_2 \setminus \kappa_1)$ is the symmetric difference of $\kappa_1$ and $\kappa_2$. In particular, the transpose of $M^{\top_\kappa}$ is $M^{\top_{\BAR\kappa}}$ where $\BAR\kappa \defeq [d] \setminus \kappa$ is the complement of $\kappa$. If $M$ is symmetric (i.e.\ $M^\top = M$), it follows that $M^{\top_\kappa} = M^{\top_{\BAR\kappa}}$.

\begin{df}[Fully symmetric matrices]\label{def:fully-symmetric}
We say that an $n^d \times n^d$ matrix $M$ is {\em fully symmetric} if $M^{\top_k} = M$ for all $k \in [d]$ (equivalently: if $M^{\top_\kappa} = M$ for all $\kappa \subseteq [d]$).
\end{df}

\begin{ex}
Let $d = 2$ and $n = 3$. Suppose $M$ is a partitioned $3^2 \times 3^2$ matrix with $3 \times 3$ blocks $M_{11},\dots,M_{33}$. Then
\[
&&
M^{\top_\emptyset} = M &= 
\small
\begin{bmatrix}
M_{11} & M_{12} & M_{13} \\
M_{21} & M_{22} & M_{23} \\
M_{31} & M_{32} & M_{33}
\end{bmatrix},
&
M^{\top_{\{1\}}} = M^{\top_1} &=
\small
\begin{bmatrix}
M_{11}^\top & M_{12}^\top & M_{13}^\top \\
M_{21}^\top & M_{22}^\top & M_{23}^\top \\
M_{31}^\top & M_{32}^\top & M_{33}^\top
\end{bmatrix},
&&\\&&
M^{\top_{\{2\}}} = M^{\top_2} &= 
\small
\begin{bmatrix}
M_{11} & M_{21} & M_{31} \\
M_{12} & M_{22} & M_{32} \\
M_{13} & M_{23} & M_{33}
\end{bmatrix},&
M^{\top_{\{1,2\}}} = M^\top &=
\small
\begin{bmatrix}
M_{11}^\top & M_{21}^\top & M_{31}^\top \\
M_{12}^\top & M_{22}^\top & M_{32}^\top \\
M_{13}^\top & M_{23}^\top & M_{33}^\top
\end{bmatrix}.
\]
When $d=2$ (as in this example), the operation $\top_1$ is also known as the {\em block transpose}, since it transposes the individual blocks within a partitioned $n^2 \times n^2$ matrix. 
\end{ex}

\begin{df}[Partial transpose rank]
Let $M$ be an $n^d \times n^d$ matrix.
\begin{itemize}
\item
$M$ is {\em PT-basic} if there exists $\kappa \subseteq [d]$ such that $\rank(M^{\top_\kappa}) = 1$.
\item
The {\em PT-rank} of $M$ is the minimum number of PT-basic matrices that sum to $M$.
\end{itemize}
\end{df}

The next lemma gives an alternative characterization of $\PTrank$.

\begin{la}\label{la:decomp1}
The PT-rank of an $n^d \times n^d$ matrix $M$ equals the minimum value of $\sum_{\kappa \subseteq [d-1]} \rank(N_\kappa^{\top_\kappa})$ over all decompositions $M = \sum_{\kappa \subseteq [d-1]} N_\kappa$.
\end{la}

\begin{proof}
Let $r$ be the PT-rank of $M$ and let $B_1,\dots,B_r$ be PT-basic matrices that sum to $M$. For each $B_i$, there exists $\kappa_i \subseteq [d]$ with $\rank(B_i^{\smash{\top_{\kappa_i}}}) = 1$.  Without loss of generality, we may choose $\kappa_i \subseteq [d-1]$, since $\rank(B_i^{\smash{\top_{\kappa_i}}}) = \rank((B_i^{\smash{\top_{\kappa_i}}})^\transpose) = \rank(B_i^{\smash{\top_{\BAR{\kappa_i}}}})$.  Now define $N_\kappa \defeq \sum_{i \in [r]\,:\,\kappa_i=\kappa} B_i$.
\end{proof}

The next two examples illustrate PT-rank in the case $d=2$.

\begin{ex}
Let $d = 2$ and $n = 2$. The $2^2 \times 2^2$ identity matrix has PT-rank $2$, since it is not PT-basic, but is a sum of PT-basic matrices:
\small
\[
\left[
\begin{array}{rr|rr}
  1 &  0  &  0 &  0  \\
  0 &  1  &  0 &  0  \\
\hline
  0  &  0 &  1 &  0  \\
  0  &  0 &  0 &  1  \\
\end{array}
\right]
&=
\left[
\begin{array}{rr|rr}
  1 &  0  &   0 &  \!\!\!-1  \\
  0 &  0  &  0 &  0  \\
\hline
  0  & 0 &  0  &  0  \\
  \!\!\!-1  &  0 &  0  &  1  \\
\end{array}
\right]
+
\left[
\begin{array}{rr|rr}
  0 &  0  &  0 &  0  \\
  0 &  1  &  1  &  0  \\
\hline
  0  & 1  &  1  &  0  \\
  0  &  0 &  0  &  0  \\
\end{array}
\right]^{\top_1}
\]
\end{ex}

\begin{ex}\label{ex:3-squared}
The $3^2 \times 3^2$ matrix on the left below is PT-basic (has PT-rank $1$), yet has rank $9$:
\small
\[
\left[
\begin{array}{ccc|ccc|ccc}
  1 & 0 & 0   &   0 & 0 & 0   &   0 & 0 & 0  \\
  0 & 0 & 0   &   1 & 0 & 0   &   0 & 0 & 0  \\
  0 & 0 & 0   &   0 & 0 & 0   &   1 & 0 & 0  \\
\hline
  0 & 1 & 0   &   0 & 0 & 0   &   0 & 0 & 0  \\
  0 & 0 & 0   &   0 & 1 & 0   &   0 & 0 & 0  \\
  0 & 0 & 0   &   0 & 0 & 0   &   0 & 1 & 0  \\
\hline
  0 & 0 & 1   &   0 & 0 & 0   &   0 & 0 & 0  \\
  0 & 0 & 0   &   0 & 0 & 1   &   0 & 0 & 0  \\
  0 & 0 & 0   &   0 & 0 & 0   &   0 & 0 & 1  
\end{array}
\right]
=
\left[
\begin{array}{ccc|ccc|ccc}
  1 & 0 & 0   &   0 & 1 & 0   &   0 & 0 & 1  \\
  0 & 0 & 0   &   0 & 0 & 0   &   0 & 0 & 0  \\
  0 & 0 & 0   &   0 & 0 & 0   &   0 & 0 & 0  \\
\hline
  0 & 0 & 0   &   0 & 0 & 0   &   0 & 0 & 0  \\
  1 & 0 & 0   &   0 & 1 & 0   &   0 & 0 & 1  \\
  0 & 0 & 0   &   0 & 0 & 0   &   0 & 0 & 0  \\
\hline
  0 & 0 & 0   &   0 & 0 & 0   &   0 & 0 & 0  \\
  0 & 0 & 0   &   0 & 0 & 0   &   0 & 0 & 0  \\
  1 & 0 & 0   &   0 & 1 & 0   &   0 & 0 & 1  
\end{array}
\right]^{\top_1}
\]
\end{ex}

\subsection{Properties of PT-rank}\label{sec:pt-props}

Unlike the full transpose ($\kappa=[d]$), the partial transpose operation can increase matrix rank when $\emptyset \subsetneqq \kappa \subsetneqq [d]$. 

\begin{la}
For every $n^d \times n^d$ matrix $M$ and $\kappa \subseteq [d]$, we have 
\[
  \rank(M^{\top_\kappa}) \le n^{2\min\{|\kappa|,d-|\kappa|\}}\, \rank(M).
\]
Moreover, this inequality is best possible in general (as illustrated by Example~\ref{ex:3-squared} in the case $d=2$ and $|\kappa|=1$).
\end{la}

\begin{proof}
We show the calculation when $\kappa=\{d\}$ and $M$ has rank $1$, and the more general case follows easily. Consider the case when $M$ has rank $1$. Then, $M=AB^{\transpose}$ for some $A,B\in \F^{n^d}$. Consider $\vec i,\vec j\in [n]^{d-1}$ and $u,v\in [n]$. We have
\[
    M_{\vec iu,\vec jv}^{\top_\kappa}
    =M_{\vec iv, \vec ju}
    =A_{\vec iv}\cdot B_{\vec ju}
    =\sum_{a,b\in [n]}(\mathbbm{1}[v=b]\cdot A_{\vec ia}) \cdot (\mathbbm{1}[u=a]\cdot B_{\vec jb}).
\]
Define $A^{(a,b)}$ by $(A^{(a,b)})_{\vec iv}=\mathbbm{1}[v=b]\cdot A_{\vec ia}$ and define $B^{(\vec a,\vec b)}$ by $(B^{(\vec a,\vec b)})_{\vec ju}=\mathbbm{1}[u=a]\cdot B_{\vec jb}$. From the above, we see that $M=\sum_{a,b\in[n]}A^{(a,b)}(B^{(a,b)})^\transpose$, so $\rank(M^{\kappa})\leq n^2$.
\end{proof}

The partial transpose also behaves differently than the full transpose with respect to products of matrices. For two $n^d \times n^d$ matrices $P$ and $M$, we have $(PM)^\top = M^\top P^\top$; additionally, $\rank((PM)^\top) = \rank(M^\top)$ whenever $P$ is invertible.
With respect to the partial transpose, we get analogous equations only in the case that $P$ (or $M$) is a Kronecker product of $d$ $n \times n$ matrices.

\begin{la}\label{la:invariance}
Let $P$ and $M$ be $n^d \times n^d$ matrices such that $P = B_1 \kron \dots \kron B_d$ for $n \times n$ matrices $B_1,\dots,B_d$. Then for all $\kappa \subseteq [d]$,
\[
  (PM)^{\top_\kappa} = LM^{\top_\kappa}R\quad
  \text{where }
  L =
  \bigkron_{k=1}^d
  \begin{cases}
    I_n &\text{if }k \in \kappa\\
    B_k &\text{otherwise}
  \end{cases}
  \text{ and }
  R = 
  \bigkron_{k=1}^d
  \begin{cases}
    B_k^\transpose &\text{if }k \in \kappa\\
    I_n &\text{otherwise.}
  \end{cases}
\]
Moreover, if $B_1,\dots,B_d$ are nonsingular then $\rank((PM)^{\top_\kappa}) = \rank(M^{\top_\kappa})$.
\end{la}

\begin{proof}
We show the calculation when $\kappa = \{d\}$. 
Here $L = B_1 \kron \dots \kron B_{d-1} \kron I_n$ and $R = I_{n^{d-1}} \kron B_d^\transpose$.
Consider $\vec i,\vec j \in [n]^{d-1}$ and $u,v \in [n]$.
We have
\[
  (PM)^{\top_d}_{\vec iu,\vec jv}
  =
  (PM)_{\vec iv,\vec ju}
  &=
  \sum_{\vec h \in [n]^{d-1}} \sum_{s \in [n]}
  P_{\vec iv,\vec hs} M_{\vec hs,\vec ju}
  \\
  &=
  \sum_{\vec h \in [n]^{d-1}} \sum_{s \in [n]}
  \Big(\prod_{c=1}^{d-1} (B_c)_{i_c,h_c}\Big) (B_d)_{v,s}
  M_{\vec hs,\vec ju}
  \\
  &=
  \sum_{\vec h \in [n]^{d-1}} \sum_{s \in [n]}
  \Big(\prod_{c=1}^{d-1} (B_c)_{i_c,h_c}\Big)
  M^{\top_d}_{\vec hu,\vec js}
  (B_d^\transpose)_{s,v}
  \\
  &=
  \sum_{\vec hr \in [n]^d} \sum_{\vec gs \in [n]^d}
  \Big(\prod_{c=1}^{d-1} (B_c)_{i_c,h_c}\Big)
  (I_n)_{u,r}
  M^{\top_d}_{\vec hr,\vec gs}
  (I_{n^{d-1}})_{\vec g,\vec j}
  (B_d^\transpose)_{s,v}
  \\
  &=
  \sum_{\vec hr \in [n]^d} \sum_{\vec gs \in [n]^d}
  L_{\vec iu,\vec hr}
  M^{\top_d}_{\vec hr,\vec gs}
  R_{\vec gs,\vec jv}
  =
  (L
  M^{\top_d}
  R)_{\vec iu,\vec jv}.\qedhere
\]
\end{proof}

\begin{cor}\label{cor:invariance}
$\PTrank$ is invariant under the action of $\mr{GL}(n)^d$ on $n^d \times n^d$ matrices.
\end{cor}

Corollary \ref{cor:invariance} shows that $\PTrank$ generalizes to a complexity measure on multilinear functions $V_1 \times \dots \times V_d \to V_1 \times \dots \times V_d$ (equivalently: on tensors in $\bigotimes_{c=1}^d (V_c \otimes V_c^\ast)$) for any $n$-dimensional $\F$-vector spaces $V_1,\dots,V_d$, independent of a choice of bases.

Note that the PT-rank of a matrix depends on its assumed dimension. For example, the PT-rank of an $n^4 \times n^4$ matrix is different from its PT-rank treated as a $\wh n^2\times \wh n^2$ matrix where $\wh n = n^2$, as a different set of partial transposes are allowed. When there is ambiguity, we write $\PTrank_{[n]^4}(M)$ or $\PTrank_{[n^2]^2}(M)$ to distinguish these two cases. 

\begin{la}
$\PTrank_{[n]^{pq}}(M) \le \PTrank_{[n^p]^q}(M)$ whenever $d = pq$ with $p,q \ge 1$.
\end{la}

\begin{proof}
Write $M$ for the $n^{pq} \times n^{pq}$ matrix and $N$ for the equivalent $\wh n^q \times \wh n^q$ matrix with $\wh n = n^p$. For $\lambda \subseteq [q]$, note that
\[
  \rank(N^{\top_\lambda}) = \rank(M^{\top_\kappa})
  \quad\text{where } \kappa = \{k + (l-1)p : k \in [p],\ l \in [\lambda]\}.
\]
The claimed inequality follows from this fact and the characterization of PT-rank given by Lemma \ref{la:decomp1}.
\end{proof}

\subsection{PT-rank and SoS complexity}\label{sec:PT-SOS} 

We now explain the close relationship between PT-rank of $n^d \times n^d$ matrices and SoS complexity of $d$-multiquadratic forms.

\begin{df}[Multiquadratic polynomial $Q_M$]
For an $n_1\cdots n_d \times n_1\cdots n_d$ matrix $M$, let $Q_M$ denote the $d$-multiquadratic polynomial
\[
  Q_M
  \defeq
  \sum_{\vec i,\vec j \in [n_1]\times\dots\times[n_d]} 
  M_{\vec i,\vec j} \cdot X^{(1)}_{i_1} X^{(1)}_{j_1} \cdots X^{(d)}_{i_d} X^{(d)}_{j_d}.
\]
For example, the identity matrix $I_{n^3}$ has corresponding $3$-multiquadratic polynomial
\[
  q_{I_{n^3}} = 
  \Big(\sum_{i \in [n]} X_i^2\Big)\Big(\sum_{i \in [n]} Y_i^2\Big)\Big(\sum_{i \in [n]} Z_i^2\Big)
\]
where we write $X_i,Y_i,Z_i$ for $X^{(1)}_i,X^{(2)}_i,X^{(3)}_i$. (The biquadratic polynomial $q_n$ in the Introduction is given by $q_{I_{n^2}}$ in this notation.)
\end{df}

\begin{la}\label{la:SOS-PT}
Let $\F$ be algebraically closed field of characteristic $\neq 2$. 
Then over $\F$, we have $\SOS(Q_M) \le \PTrank(M)$.  
\end{la}

\begin{proof}
Suppose $\SOS(Q_M)= m<\infty$. Then, there exist an $m\times n^d$ matrix $L$ such that
\[
  Q_M 
  &= 
  \sum_{\ell=1}^m \Big(\sum_{\vec i \in [n]^d} 
     L_{\ell,\vec i} \cdot X^{(1)}_{i_1} \cdots X^{(d)}_{i_d}
  \Big)^2
  =
  \sum_{\vec i,\vec j \in [n]^d} (L^\transpose L)_{\vec i,\vec j}
  \cdot X^{(1)}_{i_1} X^{(1)}_{j_1} \cdots X^{(d)}_{i_d} X^{(d)}_{j_d}.
\]

Over an algebraically closed field, we have the familiar fact
\[
  \{\tu{symmetric matrices of rank}\le m\} = \{L^\transpose L : L \in \F^{m \times n^d}\}.
\]
Therefore,
\[
  \SOS(Q_M)= 
  \min\bigg\{\rank(D) : \tu{symmetric } D \textup{ s.t.\ }
  Q_M=\sum_{\vec i,\vec j \in [n]^d} D_{\vec i,\vec j}
  \cdot X^{(1)}_{i_1} X^{(1)}_{j_1} \cdots X^{(d)}_{i_d} X^{(d)}_{j_d}\bigg\}.
\]
Since variables commute, the righthand equation is equivalent to
$M = \frac{1}{2^d} \sum_{\kappa \subseteq [d]} D^{\top_\kappa}$.
That is,
\[
  \SOS(Q_M)
  \leq 
  \min\bigg\{\rank(D) :
  M = \frac{1}{2^d} \sum_{\kappa \subseteq [d-1]} D^{\top_\kappa}\bigg\}.
\]
Given any decomposition $M = \sum_{\kappa \subseteq [d]} P_\kappa^{\top_\kappa}$ with $\PTrank(M) = \sum_\kappa \rank(P_\kappa)$, let $D = \frac{1}{2^d} \sum_\kappa P_\kappa$. Then
\[
  \frac{1}{2^d} \sum_\kappa D^{\top_\kappa}
  =
  \frac{1}{2^d} \sum_\kappa \sum_\lambda P_\lambda^{\top_\kappa}
  =
  \frac{1}{2^d} \sum_\kappa (\sum_\lambda P_\lambda^{\top_\lambda})^{\top_\kappa}
  =
  \frac{1}{2^d} \sum_\kappa M^{\top_\kappa}
  =
  M.
\]
Therefore, $\SOS(Q_M)\le \rank(D) \le \sum_\kappa \rank(P_\kappa) = \PTrank(M)$.
\end{proof}

\begin{rmk}
    Over finite fields $\F$, the bound $\SOS(Q_M)-1\leq \PTrank(M)$ can be proved using same ideas as above. The only difference is that following the result from \cite{seroussi1980factorization}, every symmetric matrix of rank $m$ can be expressed as $L^\transpose L$ for $L\in \F^{(m+1)\times n^d}$, thus comes the additional $1$. 
\end{rmk}

As we show next, the bounds above are tight up to $O(2^d)$ factor for fully symmetric matrices.

\begin{la}\label{la:PT-SOS}
If $M$ is a fully symmetric matrix over a field of characteristic $\neq 2$, then $\PTrank(M) \le 2^{d-1}\,\SOS(Q_M)$.
\end{la}

\begin{proof}
Suppose $\SOS(Q_M) = m < \infty$. Then there exists an $m \times n^d$ matrix $L$ such that
\[
  Q_M 
  &= 
  \sum_{\ell=1}^m \Big(\sum_{\vec i \in [n]^d} 
     L_{\ell,\vec i} \cdot X^{(1)}_{i_1} \cdots X^{(d)}_{i_d}
  \Big)^2
  =
  \sum_{\vec i,\vec j \in [n]^d} (L^\transpose L)_{\vec i,\vec j}
  \cdot X^{(1)}_{i_1} X^{(1)}_{j_1} \cdots X^{(d)}_{i_d} X^{(d)}_{j_d}.
\]
Since these are commutative polynomials (in particular, $X^{(k)}_{i_k} X^{(k)}_{j_k} = X^{(k)}_{j_k} X^{(1)}_{i_k}$), it follows that
\[
  \sum_{\kappa \subseteq [d]} M^{\top_\kappa} = \sum_{\kappa \subseteq [d]} (L^\transpose L)^{\top_\kappa}.
\]
Since $M$ is fully symmetric and $L^\transpose L$ is symmetric, it follows that $2^{d-1}\, M = \sum_{\kappa \subseteq [d-1]} (L^\transpose L)^{\top_\kappa}$. Therefore, we have $M = \sum_{\kappa \subseteq [d-1]} N_\kappa$ for matrices $N_\kappa \defeq \frac{1}{2^{d-1}} (L^\transpose L)^{\top_\kappa}$. 
Since $\rank((N_\kappa)^{\top_\kappa}) = \rank(L^\transpose L) = s$ for each $\kappa \subseteq [d-1]$, we have $\PTrank(M) \le 2^{d-1} s = 2^{d-1}\,\SOS(Q_M)$.
\end{proof}

Lemmas \ref{la:PT-SOS} and \ref{la:SOS-PT} together imply:

\begin{prop}\label{prop:pt-rank-bil}
Let $M$ be an $n^d \times n^d$ matrix over algebraically closed or finite field of characteristic $\neq 2$. Then $\SOS(Q_M) - 1 \le \PTrank(M) \le 2^{d-1}\,\SOS(Q_M)$.
\end{prop}

\subsection{PT-rank of almost all matrices}

Although not required for any other results in this paper, we conclude this section by observing that (asymptotically) almost all $n^d \times n^d$ matrices have PT-rank $\Omega(n^d)$, and (asymptotically) almost all fully symmetric $n^d \times n^d$ matrices have PT-rank $\Omega((n/2)^d)$.  These results are established standard dimension-counting arguments.

\begin{prop}\label{prop:almost-all}
Asymptotically almost all $n^d \times n^d$ matrices have PT-rank at least $(\frac{1}{2}-\eps) n^d$ for every constant $\eps>0$.
\end{prop}

\begin{proof}
First consider the case that $\F$ is a finite field. There are $|\F|^{n^{2d}}$ distinct $n^d \times n^d$. 
At most $|\F|^{2n^d}$ of these have rank $1$ (since 
every rank-1 matrix is an outer product of two $n^d$-dimensional vectors), and therefore at most $|\F|^{2rn^d}$ of these have rank $\leq r$. 

We now count the number of matrices with PT-rank $\leq r$. A matrix $M$ of PT-rank at most $r$ can be expressed as $\sum_{\kappa\subseteq [d]}M_{\kappa}^{\top_\kappa}$, where $r\geq \sum_{\kappa\subseteq [d]}r_{\kappa}$ for $r_{\kappa}=\rank(M_{\kappa})$. For a fixed set of $\{r_\kappa\}$ with $r= \sum_{\kappa\subseteq [d]}r_{\kappa}$, the number of $\{M_\kappa\}$ is $\prod_{\kappa} |\F|^{2r_{\kappa}n^d}=|\F|^{2rn^d}$ (note that this count also includes $\{M_{\kappa}'\}$ for which $\{r_\kappa'\}$ is strictly dominated by $\{r_\kappa\}$). The number of such $\{r_\kappa\}$ is at most $\binom{r+2^d}{2^d}$, counted with a balls and urns argument, which is less than $(r+2^d)^{2^d}$. Therefore, the number of matrices with PT-rank $\leq r$ is bounded by $(r+2^d)^{2^d}\cdot |\F|^{2rn^d}$. Therefore, for a uniform random $n^d \times n^d$ matrix $M$, we have
\[
\Pr\big[\ 
    \text{$M$ has PT-rank $\le$ }(\textstyle\frac{1}{2}-\varepsilon)n^d\big]\leq \frac{((\frac{1}{2}-\eps)n^d+2^d)^{2^d}\cdot |\F|^{2((\frac{1}{2}-\eps)n^d)n^d}}{|\F|^{n^{2d}}}\leq \frac{n^{d\cdot 2^d}}{|\F|^{2\eps n^d}}=o(1).
\]

If $\F$ is an infinite field, then a similar dimension-counting argument shows that the Zariski closure of the set of $n^d \times n^d$ matrices with PT-rank at most $(\frac{1}{2}-\eps)n^d$ has dimension strictly less than $n^{2d}$.
\end{proof}

Since the set of fully symmetric $n^d \times n^d$ matrices has dimension $\ge n^{2d}/2^d$, a similar counting argument shows:

\begin{prop}\label{prop:fully-symmetric}
Asymptotically almost all  fully symmetric $n^d \times n^d$ matrices have PT-rank at least $(\frac{1}{2^{d+1}}-\eps) n^d$ for any constant $\eps>0$.
\end{prop}

\section{Non-commutative set-multilinear formula lower bounds from PT-rank}\label{sec:warm-up}

Recall the main theorem of this paper:

\begin{thmrest}
For every $n^d \times n^d$ matrix $M$, we have
\[
  \Lsm(\ShiftT{M}) 
  &\ge
  \frac{\PTrank(M)}{n^{d-\log d+1}}.
\]
\end{thmrest}

As a helpful warm-up to the proof in \S\ref{sec:proof}, in this section we shall prove a strictly weaker---but slightly simpler---version of Theorem \ref{thm:the-main} that lower bounds the \emph{non-commutative} set-multilinear formula size of $\ShiftT{M}$.  We start with the definition:

\begin{df}
The {\em non-commutative set-multilinear formula size} of a tensor $A : [n]^d \to \F$, denoted $\Lnc(A)$, is defined inductively as follows:
\begin{itemize}
  \item
    If $d = 1$, then $\Lnc(A)$ equals $1$ if $A$ is non-zero at any point and $0$ otherwise.
  \item
    If $d \ge 2$, then $\Lnc(A)$ is the minimum value of $\sum_i\ (\Lnc(B_i) + \Lnc(C_i))$
    over indexed families $\{(e_i,f_i,B_i,C_i)\}_i$ where $e_i,f_i$ are positive integers satisfying $e_i + f_i = d$ and $B_i : [n]^{e_i} \to \F$ and $C_i : [n]^{f_i} \to \F$ are tensors such that $A = \sum_i B_i \otimes C_i$.
\end{itemize}
\end{df}

In the remainder of this section, we prove the following:
\begin{thm}[Non-commutative version of Theorem \ref{thm:the-main}]\label{thm:nc-main}
For every $n^d \times n^d$ matrix $M$, we have
\[
  \Lnc(\ShiftT{M}) 
  &\ge
  \frac{\PTrank(M)}{n^{d-\log d+1}}.
\]
\end{thm}

\subsection{The complexity measure $\rho$ on order $2d$ tensors}

We introduce a complexity measure on tensors $A : [n]^{2d} \to \F$, denoted $\rho(A)$, which is closely related to PT-rank (normalized by factor $n^{-d}$), but with an added ``max'' quantifier that affects the $\top_1$ and $\top_k$ partial transposes.  (A precise relationship between $\rho$ and $\PTrank$ is noted at the end of this section.)

\begin{df}
For any $a,b \in \{0,1\}$ and $\alpha \in \{0,1\}^{d-1}$, we define a partition $I_{a,\alpha,b} \sqcup J_{a,\alpha,b} = [2d]$ by
\begin{alignat*}{4}
  I_{a,\alpha,b}
  &\defeq
  [2d] \,\cap\, \{&&a,\,2+\alpha_1,\,4+\alpha_2,\,\dots,\,2d-2+\alpha_d,\,2d+b&&\},\\
  J_{a,\alpha,b}
  &\defeq
  [2d] \,\cap\, \{1-\mbox{}&&a,\,3-\alpha_1,\,5-\alpha_2,\,\dots,\,2d-1-\alpha_d,\,2d+1-b&&\}.
\end{alignat*}

For a tensor $A : [n]^{2d} \to \F$, we define its {\em relative rank} with respect to $(a,\alpha,b) \in \{0,1\}^{d+1}$ as follows:
\[
  \relrk_{a,\alpha,b}(A)
  &\defeq
  n^{-d}\ \rank(\Mat_{I_{a,\alpha,b},J_{a,\alpha,b}}(A)).
\]
\end{df}

\begin{la}\label{la:deficit}
$\relrk_{a,\alpha,b}(A) 
  \le
  n^{-\mathds{1}[a\ne b]}$.
\end{la}

\begin{proof}
Follows from the observation that
\[
  |I_{a,\alpha,b}| &= d-1 + \mathds 1[a=1] + \mathds 1[b=0],\\
  |J_{a,\alpha,b}| &= d-1 + \mathds 1[a=0] + \mathds 1[b=1],
\]
and therefore $\min\{|I_{a,\alpha,b}|,\ |J_{a,\alpha,b}|\} = d-\mathds 1[a\ne b]$.
\end{proof}

The next lemmas show that $\relrk$ is sub-additive and multiplicative. 

\begin{la}\label{la:relrk-subadd}
$\relrk_{a,\alpha,b}(\sum_i A_i) \le \sum_i\ \relrk_{a,\alpha,b}(A_i)$.
\end{la}

\begin{proof}
Follows from sub-additivity of rank.
\end{proof}

\begin{la}\label{la:relrk-mult}
For all $A : [n]^{2d} \to \F$ and $B : [n]^{2e} \to \F$ and $a,b,c \in \{0,1\}$ and $\alpha \in \{0,1\}^{d-1}$ and $\beta \in \{0,1\}^{e-1}$, with respect to the tensor product $A \otimes B : [n]^{2(d+e)} \to \F$, we have
\[
  I_{a,\alpha,b,\beta,c}
  &=
  I_{a,\alpha,b} \sqcup I_{b,\beta,c},\\
  J_{a,\alpha,b,\beta,c}
  &=
  J_{a,\alpha,b} \sqcup J_{b,\beta,c},\\
  \Mat_{I_{a,\alpha,b,\beta,c},J_{a,\alpha,b,\beta,c}}(A \otimes B)
  &=
  \Mat_{I_{a,\alpha,b},J_{a,\alpha,b}}(A)
  \kron
  \Mat_{I_{b,\beta,c},J_{b,\beta,c}}(B),\\
  \relrk_{a,\alpha,b,\beta,c}(A \otimes B)
  &=
  \relrk_{a,\alpha,b}(A)\
  \relrk_{b,\beta,c}(B).
\]
\textup{(Note: We only require the inequality $\relrk_{a,\alpha,b,\beta,c}(A \otimes B) \le
  \relrk_{a,\alpha,b}(A)\ 
  \relrk_{b,\beta,c}(B)$, that is, 
the sub-multiplicativity of $\relrk$.)}
\end{la}

\begin{proof}
Follows from definitions and multiplicativity of matrix rank under Kronecker product.
\end{proof}

\begin{df}
For a tensor $A : [n]^{2d} \to \F$, notation $\{X_\alpha\} \vdash A$ denotes that $X_\alpha$ are tensors $[n]^{2d} \to \F$, indexed by $\alpha \in \{0,1\}^{d-1}$, such that $A = \sum_\alpha X_\alpha$.

We define a complexity measure $\rho(A)$ by
\[
  \rho(A)
  &\defeq
  \max_{a,b \in \{0,1\}}\ 
  \min_{\{X_\alpha\} \vdash A}\ 
  \sum_{\alpha \in \{0,1\}^{d-1}}\ 
  \relrk_{a,\alpha,b}(X_\alpha).
\]
\end{df}

\begin{la}\label{la:rho-subadd}
$\rho$ is sub-additive, that is, $\rho(\sum_i A_i) \le \sum_i \rho(A_i)$ where $\{A_i\}_i$ is any indexed family of tensors $A_i : [n]^{2d} \to \F$.
\end{la}

\begin{proof}
We have
\[
  \rho(\sum_i A_i)
  &=
  \max_{a,b}\ 
  \:\hspace{1pt}
  \min_{\{X_\alpha\} \vdash \sum_i A_i}\
  \:\hspace{1pt}
  \sum_\alpha\ 
  \relrk_{a,\alpha,b}(X_\alpha)\\
  &\le
  \max_{a,b}\ 
  \min_{\substack{
    \{X_{i,\alpha}\} \vdash A_i\\
    X_\alpha \,\defeq\, \sum_i X_{i,\alpha}
  }}\
  \sum_\alpha\ 
  \relrk_{a,\alpha,b}(X_\alpha)\\
  &=
  \max_{a,b}\ 
  \min_{\{X_{i,\alpha}\} \vdash A_i}\
  \sum_\alpha\ 
  \relrk_{a,\alpha,b}(\sum_i X_{i,\alpha})\\
  &\le
  \max_{a,b}\ 
  \min_{\{X_{i,\alpha}\} \vdash A_i}\
  \sum_i\ 
  \sum_\alpha\ 
  \relrk_{a,\alpha,b}(X_{i,\alpha})\\
  &=
  \max_{a,b}\ 
  \sum_i\ 
  \min_{\{X_{i,\alpha}\} \vdash A_i}\
  \sum_\alpha\ 
  \relrk_{a,\alpha,b}(X_{i,\alpha})\\
  &\le
  \sum_i\ 
  \max_{a,b}\ 
  \min_{\{X_{i,\alpha}\} \vdash A_i}\
  \sum_\alpha\ 
  \relrk_{a,\alpha,b}(X_{i,\alpha})\\
  &=
  \sum_i\ \rho(A_i).\qedhere
\]
\end{proof}

Unlike relative rank, $\rho$ is not obviously multiplicative or sub-multiplicative under tensor product.  However, the following key lemma shows that $\rho$ {\em always} shrinks under tensor products.  In contrast, the relative rank of a tensor product $A \otimes B$ is only guaranteed to shrink when (the matrix flattening of) $A$ or $B$ has unbalanced dimensions.

\begin{la}\label{la:rho-tensor}
For all $A : [n]^{2d} \to \F$ and $B : [n]^{2e} \to \F$, we have 
\[
  \rho(A \otimes B) \le 
  n^{-1}\ \min\{\rho(A),\ \rho(B)\}.
\]
\end{la}

\begin{proof}
Let $a,b,c$ range over $\{0,1\}$,
and let $\alpha,\beta$ range over $\{0,1\}^{d-1},\{0,1\}^{e-1}$ respectively. 
We have
\begin{align*}
  \rho(A \otimes B)
  &=
  \max_{a,c}\ 
  \qquad
  \min_{\substack{
    \{Z_{\alpha,b,\beta}\} \vdash A \otimes B
  }}\
  \qquad
  \sum_{\alpha,b,\beta}\ 
  \relrk_{a,\alpha,b,\beta,c}(Z_{\alpha,b,\beta})
  \\
  &\le
  \max_{a,c}\ 
  \min_{\substack{
    \{X_\alpha\} \vdash A,\ \{Y_\beta\} \vdash B\\
    Z_{\alpha,b,\beta} \,\defeq\, \mathds 1[b = 1-c] \cdot X_\alpha \otimes Y_\beta
  }}\
  \sum_{\alpha,b,\beta}\ 
  \relrk_{a,\alpha,b,\beta,c}(Z_{\alpha,b,\beta})
  \\
  &=
  \max_{a,c}\ 
  \min_{\{X_\alpha\} \vdash A,\ \{Y_\beta\} \vdash B}\
  \sum_{\alpha,\beta}\ 
  \relrk_{a,\alpha,1-c,\beta,c}(X_\alpha \otimes Y_\beta)
  \\
  &= 
  \max_{a,c}\ 
  \min_{\{X_\alpha\} \vdash A,\ \{Y_\beta\} \vdash B}\
  \sum_{\alpha,\beta}\ 
  \relrk_{a,\alpha,1-c}(X_\alpha)\ 
  \relrk_{1-c,\beta,c}(Y_\beta)
  \quad\text{(Lemma \ref{la:relrk-mult})}
  \\
  &\le
  \max_{a,c}\ 
  \min_{\{X_\alpha\} \vdash A,\ \{Y_\beta\} \vdash B}\
  \sum_{\alpha,\beta}\ 
  \relrk_{a,\alpha,1-c}(X_\alpha)\
  n^{-1}
  \quad\text{(Lemma \ref{la:deficit})}
  \\
  &=
  n^{-1}\  
  \max_{a,c}\ 
  \min_{\{X_\alpha\} \vdash A}\
  \sum_{\alpha}\ 
  \relrk_{a,\alpha,1-c}(X_\alpha)
  \\
  &=
  n^{-1}\  
  \max_{a,b}\ 
  \min_{\{X_\alpha\} \vdash A}\
  \sum_{\alpha}\ 
  \relrk_{a,\alpha,b}(X_\alpha)
  \\
  &=
  n^{-1}\ 
  \rho(A).\qedhere
\end{align*}
The proof of $\rho(A\otimes B)\le\ n^{-1}\rho(B)$ follows similarly. 
\end{proof}

\subsection{Non-commutative formula lower bounds from $\rho$}

\begin{df}[{The order $d$ flattening $A^\flat : [n^2]^d \to \F$ of an order $2d$ tensor $A : [n]^{2d} \to \F$}]\label{df:flat1}
For an order $2d$ tensor $A : [n]^{2d} \to \F$, we define its order $d$ flattening $A^\flat : [n^2]^d \to \F$ by the identity
\[
  A(a_1,b_1,\dots,a_d,b_d) 
  =
  A^\flat(\sq{a_1,b_1},\dots,\sq{a_d,b_d})
\]
for all $a_1,b_1,\dots,a_d,b_d \in [n]$.
\end{df}

Note that this flattening operation commutes with the tensor product: for any $A : [n]^{2d} \to \F$ and $B : [n]^{2e} \to \F$, we have $(A \otimes B)^\flat = A^\flat \otimes B^\flat$ (as tensors $[n]^{2(d+e)} \to \F$).

\begin{thm}\label{thm:nc-lower-bound}
For every $A : [n]^{2d} \to \F$, we have $\Lnc(A^\flat) \ge n^{\log d}\ \rho(A)$. That is, any non-commutative set-multilinear formula computing the flattened tensor $A^\flat : [n^2]^d \to \F$ has size at least $n^{\log d}\ \rho(A)$.
\end{thm}

\begin{proof}
We argue by induction on $d$. The base case $d = 1$ is trivial. Assume $d \ge 2$ and fix $\{(e_i,f_i,B_i,C_i)\}_i$ with $e_i,f_i \ge 1$ and $e_i+f_i = d$ and $B_i : [n]^{e_i} \to \F$ and $C_i : [n]^{f_i} \to \F$ such that $A = \sum_i\ B_i \otimes C_i$ and
$
  \Lnc(A) = \sum_i\ (\Lnc(B_i) + \Lnc(C_i)).
$

We have
\[
  \Lnc(A^\flat) 
  &\ge
  \sum_i\,
  \Big(
     n^{\log e_i}\ \rho(B_i) + n^{\log f_i}\ \rho(C_i)
  \Big)
  \quad\text{(induction hypothesis)}
  \\
  &\ge
  \sum_{i\,:\, e_i \,\ge\, d/2}
  n^{\log e_i}\ \rho(B_i)
  +
  \sum_{i \,:\, f_i \,>\, d/2}
  n^{\log f_i}\ \rho(C_i)
  \\
  &\ge
  \sum_{i\,:\, e_i \,\ge\, d/2} 
  n^{\log e_i + 1}\ 
  \rho(B_i \otimes C_i)
  +
  \sum_{i\,:\, f_i \,>\, d/2}
  n^{\log f_i + 1}\ 
  \rho(B_i \otimes C_i)
  \quad\text{(Lemma \ref{la:rho-tensor})}
  \\
  &\ge
  n^{\log d}\ 
  \sum_i\,
  \rho(B_i \otimes C_i)\\
  &\ge
  n^{\log d}\ 
  \rho(
  \sum_i B_i \otimes C_i)
  \quad\text{(Lemma \ref{la:rho-subadd})}
  \\
  &=
  n^{\log d}\ \rho(A).\qedhere
\]
\end{proof}

We are finally ready to prove Theorem \ref{thm:nc-main}.

\begin{proof}[Proof of Theorem \ref{thm:nc-main}]
Let $M$ be an $n^d \times n^d$ matrix. Recall the order $d+1$ tensor $\ShiftT{M} : [n^2]^{d+1} \to \F$, which equals $\PaddedT{M}^\flat$ for the order $2d+2$ tensor $\PaddedT{M} : [n]^{2d+2} \to \F$ defined by
\begin{align*}
\PaddedT{M}(p,i_1,j_1,i_2,\dots,j_{d-1},i_d,j_d,q)
  &= \mathds 1\{p=q=1\} \cdot
  M_{(i_1,\dots,i_d),(j_1,\dots,j_d)} ,\\
  \ShiftT{M}(\sq{p,i_1},\sq{j_1,i_2},\dots,\sq{j_{d-1},i_d},\sq{j_d,q}) 
  &=
  \mathds 1\{p=q=1\} \cdot
  M_{(i_1,\dots,i_d),(j_1,\dots,j_d)}.
\end{align*}
It follows immediately from definitions of $\PTrank$ and $\rho$ that 
$\PTrank(M) = n^{d+1}\,\rho(\PaddedT{M})$.
Theorem  \ref{thm:nc-lower-bound} now implies the desired bound
\[\Lnc(\ShiftT{M}) 
  \ge
  n^{\log d}\ \rho(\PaddedT{M})
  &=
  \frac{\PTrank(M)}{n^{d-\log d+1}}.\qedhere
\]
\end{proof}

\section{Commutative set-multilinear formula lower bounds from PT-rank}\label{sec:proof}

This section generalizes the argument in the last section to prove our main theorem (Theorem \ref{thm:the-main}).

\subsection{Extending $\rho$ to tensors over subgraphs of $\mr{Path}_d$}

Instead of considering tensors $[n]^{2d} \to \F$, we consider tensors $[n]^{D(G)} \to \F$ with coordinates indexed by the (symmetric) directed edge set of a subgraph $G$ of an undirected path graph.

Let $\mr{Path}_\Z$ be the undirected path graph with vertex set $\{v_i : i \in \Z\}$ and edge set $\{\{v_{i-1},v_i\} : i \in \Z\}$.

Let $G = (V(G),E(G))$, $E(G) \subseteq \binom{V(G)}{2}$, range over nonempty finite subgraphs of $\mr{Path}_\Z$ with no isolated vertices (i.e.,\ $V(G) = \bigcup_{e \in E(G)} e$). 
We next introduce some useful notation:
\begin{itemize}
  \item
    We consider the partition $V_1(G) \sqcup V_2(G) = V(G)$ defined by
    \[
      V_1(G) &\defeq \{v \in V(G) : v \text{ has degree $1$ in } G\},\\
      V_2(G) &\defeq \{v \in V(G) : v \text{ has degree $2$ in } G\}.
    \]
  \item
    Let $D(G) \subseteq V(G) \times V(G)$ be the (symmetric) set of directed edges of $G$, that is,
    \[
      D(G) &\defeq \{vw : \{v,w\} \in E(G)\}.
    \] 
    We consider the partition $D_1(G) \sqcup D_2(G) = D(G)$ defined by
    \[
      D_1(G) &\defeq \{vw \in D(G) : v \in V_1(G)\},\\
      D_2(G) &\defeq \{vw \in D(G) : v \in V_2(G)\}.
    \]
    Note that $|D_1(G)| = |V_1(G)|$ and $|D_2(G)| = 2|V_2(G)|$.
\end{itemize}

\begin{itemize}
\item
For any set $S \subseteq V(\mr{Path}_\Z)$, let $\mc P(S)$ denote the set of pairs $\gamma = (I_\gamma,J_\gamma)$ of disjoint sets $I_\gamma,J_\gamma \subseteq D(\mr{Path}_\Z)$ such that 
\[
  S = \{v : vu \in I_\gamma\} = \{v : vw \in J_\gamma\}.
\]
(That is, for each $v_i \in S$ with neighbors $v_{i-1},v_{v_{i+1}}$ in $\mr{Path}_\Z$, exactly one of the sets $I_\gamma$ and $J_\gamma$ contains $v_iv_{i-1}$ and the other contains $v_iv_{v_{i+1}}$.)
Note that $|I_\gamma|=|J_\gamma|=|S|$ and $|\mc P(S)| = 2^{|S|}$.
\item
For disjoint sets $S_1,S_2 \subseteq V(\mr{Path}_\Z)$ and $\gamma_1 \in \mc P(S_1)$ and $\gamma_2 \in \mc P(\gamma_2)$, let 
\[
\gamma_1 \sqcup \gamma_2 \defeq (I_{\gamma_1} \sqcup I_{\gamma_2},J_{\gamma_1} \sqcup J_{\gamma_2}) \in \mc P(S_1 \sqcup S_2).
\]
\end{itemize}

We now consider tensors $A : [n]^{D(G)} \to \F$ and define analogues of $\relrk$ and $\rho$. 

\begin{rmk}
When $G$ is a path of length $d$ with vertices $v_0,\dots,v_d$, we identify the sequence of directed edges $v_0v_1,v_1v_0,v_1v_2,v_2v_1,\dots,v_{d-1}v_d,v_dv_{d-1}$ with integers $1,\dots,2d$. 
We thus identify $D(G)$ with the set $[2d]$ and tensors $[n]^{D(G)} \to \F$ with tensors $[n]^{2d} \to \F$.
Under this identification, the following definitions of $\relrk$ and $\rho$ directly generalize those in \S\ref{sec:warm-up}.
\end{rmk}

\begin{df}
For a graph $G \subseteq \mr{Path}_\Z$ and tensor $A : [n]^{D(G)} \to \F$ and $\alpha \in \mc P(V_2(G))$ and $\gamma \in \mc P(V_1(G))$, we define
\[
  \relrk_{\alpha,\gamma}(A)
  &\defeq
  n^{-|E(G)|}\ \rank(\Mat_{
  I_\alpha \sqcup (I_\gamma \cap D_1(G)),\,
  J_\alpha \sqcup (J_\gamma \cap D_1(G))  
  }(A)).
\]
\end{df}

\begin{df}
For a graph $G \subseteq \mr{Path}_\Z$ and tensor $A : [n]^{D(G)} \to \F$, notation $\{X_\alpha\} \vdash A$ denotes that $X_\alpha$ are tensors $[n]^{D(G)} \to \F$, indexed by $\alpha \in \mc P(V_2(G))$, such that $A = \sum_\alpha X_\alpha$.

We define the complexity measure $\rho(A)$ by
\[
  \rho(A)
  &\defeq
  \max_{\gamma \in \mc P(V_1(G))}\ 
  \min_{\{X_\alpha\}\vdash A}\ 
  \sum_{\alpha \in \mc P(V_2(G))}\ 
  \relrk_{\alpha,\gamma}(A).
\]
\end{df}

\begin{la}\label{la:rho-subadd2}\label{la:relrk-subadd2}
$\rho(\sum_i A_i) \le \sum_i\rho(A_i)$.\qed
\end{la}

Note that $|I_\alpha| = |J_\alpha| = |V_2(G)|$ and $|D_1(G) \cap I_\gamma| + |D_1(G)\cap J_\gamma| = |V_1(G)|$. 
As a consequence, we get the following generalization of Lemma \ref{la:deficit}.

\begin{la}\label{la:deficit2}
$\relrk_{\alpha,\gamma}(A)
\le
n^{-|\, |I_\gamma \,\cap\, D_1(G)| \,-\, |J_\gamma \,\cap\, D_1(G)|\, |}$.\qed
\end{la}

Part (a) of the next lemma shows that the bound of Lemma \ref{la:deficit2} may be as small as $n^{-|V_1(G)|}$. Part (b) makes a related observation that will be useful later.

\begin{la}\label{la:delta-pm}
\
\begin{enumerate}[\normalfont\quad(a)]
\item
For every graph $G \subseteq \mr{Path}_\Z$, there exist unique elements $\gamma^+,\gamma^- \in \mc P(V_1(G))$ such that 
\[
  I_{\gamma^+} = J_{\gamma^-} = D_1(G),\qquad 
  J_{\gamma^+} \cap D_1(G) = I_{\gamma^-} \cap D_1(G) = \emptyset.
\]
\item
For all edge-disjoint graphs $G,H \subseteq \mr{Path}_\Z$, there exist unique elements $\delta^+,\delta^- \in \mc P(V_1(G) \cap V_1(H))$ such that
\[
  I_{\delta^+} = J_{\delta^-} = 
  D_1(G) \cap D_1(H),\qquad 
  J_{\delta^+} \cap D_1(H) = I_{\delta^-} \cap D_1(H) = \emptyset.
\]
\end{enumerate}
\end{la}

\begin{proof}
(a) Consider any $v \in V_1(G)$ with neighbors $u,w$ in $\mr{Path}_\Z$. Without loss of generality, $\{v,u\} \notin E(G)$ and $\{v,w\} \in E(G)$. We include $vw$ in $I_{\gamma^+}$ ($= J_{\gamma^-}$) and $vu$ in $I_{\gamma^-}$ ($= J_{\gamma^+}$).

(b) Consider any $v \in V_1(G) \cap V_1(H)$ with neighbors $u,w$ in $\mr{Path}_\Z$. Without loss of generality, $\{v,u\} \in E(G) \setminus E(H)$ and $\{v,w\} \in E(H) \setminus E(G)$.  
We include $vw$ in $I_{\delta^+}$ ($= J_{\delta^-}$) and $vu$ in $I_{\delta^-}$ ($= J_{\delta^+}$).
\end{proof}

\begin{la}\label{la:relrk-mult2}
Suppose that $G,H$ are edge-disjoint finite subgraphs of $\mr{Path}_\Z$, let $A : [n]^{D(G)} \to \F$ and $B : [n]^{D(H)} \to \F$, and consider the tensor product $A \otimes B : [n]^{D(G \cup H)} \to \F$.
For all
\begin{gather*}
\alpha \in \mc P(V_2(G)),\quad 
\beta \in \mc P(V_2(H)),\quad
\delta \in \mc P(V_1(G) \cap V_1(H)),\\
\xi \in \mc P(V_1(G) \setminus V_1(H)),\quad
\zeta \in \mc P(V_1(H) \setminus V_1(G)),
\end{gather*}
we have
\[
  \relrk_{\alpha \cup \beta \cup \delta,\,\xi \cup \zeta}(A \otimes B)
  &=
  \relrk_{\alpha,\, \xi \cup \delta}(A)\
  \relrk_{\beta,\, \zeta \cup \delta}(B).
\]
\end{la}

\begin{la}\label{la:rho-tensor2}
For all edge-disjoint $G,H \subseteq \mr{Path}_\Z$ and tensors $A : [n]^{D(G)} \to \F$ and $B : [n]^{D(H)} \to \F$, we have
\[
\rho(A \otimes B)
\le
n^{-|V_1(G) \,\cap\, V_1(H)|}\ 
\min\{\rho(A),\ \rho(B)\}.
\]
\end{la}

\begin{proof}
We will show just the inequality $\rho(A \otimes B) \le n^{-|V_1(G) \,\cap\, V_1(H)|}\ \rho(A)$. (The argument for the other inequality is symmetric.)

As in Lemma \ref{la:relrk-mult2}, let $\alpha,\beta,\xi,\zeta,\delta$ range over
\begin{gather*}
\alpha \in \mc P(V_2(G)),\quad 
\beta \in \mc P(V_2(H)),\quad
\delta \in \mc P(V_1(G) \cap V_1(H)),\\
\xi \in \mc P(V_1(G) \setminus V_1(H)),\quad
\zeta \in \mc P(V_1(H) \setminus V_1(G)).
\end{gather*}
Unions $\alpha \cup \beta \cup \delta$ and $\xi \cup \zeta$ are in one-to-one correspondence with elements of $\mc P(V_2(G \cup H))$ and $\mc P(V_1(G \cup H))$, respectively.

By Lemma \ref{la:delta-pm}, there exist unique elements $\delta^+,\delta^- \in \mc P(V_1(G) \cap V_1(H))$ such that
\[
  I_{\delta^+} = J_{\delta^-} = D_1(G) \cap D_1(H),\qquad 
  J_{\delta^+} \cap D_1(H) = I_{\delta^-} \cap D_1(H) = \emptyset.
\]
For each $\zeta \in \mc P(V_1(H) \setminus V_1(G))$, we define $\delta^\pm(\zeta) \in \{\delta^+,\delta^-\}$ by
\[
  \delta^\pm(\zeta) 
  \defeq
  \begin{cases}
    \delta^+ &\text{if } |I_\zeta \cap D_1(H)| \ge |J_\zeta \cap D_1(H)|,\\
    \delta^- &\text{otherwise.}
  \end{cases}
\]
Note that
\[
  \big|\ |(I_\zeta \sqcup I_{\smash{\delta^\pm(\zeta)}}) \cap D_1(H)| 
       - 
       |(&J_\zeta \sqcup J_{\smash{\delta^\pm(\zeta)}}) \cap D_1(H)|\ \big|\\
  &=
  |D_1(G) \cap D_1(H)| +
  \big|\ 
  |I_\zeta \cap D_1(H)| - |J_\zeta \cap D_1(H)|
  \ \big|\\
  &\ge
  |V_1(G) \cap V_1(H)|.\vphantom{\big|}
\]
For all $\beta$ and $\zeta$ and $Y : [n]^{D(H)} \to \F$, it follows from Lemma \ref{la:deficit2} that
\[
  \relrk_{\beta,\,\delta^\pm(\zeta) \cup \zeta}(Y)
  &\le
  n^{-|V_1(G) \,\cap\, V_1(H)|}.
\]

We now have
\[
  \rho(A \otimes B)
  &=
  \max_{\xi , \zeta}\ 
  \qquad\ 
  \min_{\{Z_{\alpha \cup \beta \cup \delta}\} \vdash A \otimes B}\ 
  \qquad\ 
  \sum_{\alpha , \beta, \delta}\ 
  \relrk_{\alpha \cup \beta \cup \delta,\,\xi \cup \zeta}(Z_{\alpha \cup \beta \cup \delta})
  \\
  &\le
  \max_{\xi,\zeta}\ 
  \min_{\substack{
    \{X_\alpha\} \vdash A,\ \{Y_\beta\} \vdash B\\
    Z_{\alpha \cup \beta \cup \delta} \,\defeq\,
    \mathds 1[\delta=\delta^\pm(\zeta)] \cdot X_\alpha \otimes Y_\beta\\
  }}\ 
  \sum_{\alpha,\beta,\delta}\ 
  \relrk_{\alpha \cup \beta \cup \delta,\,\xi \cup \zeta}(Z_{\alpha \cup \beta \cup \delta})
  \\
  &=
  \max_{\xi,\zeta}\ 
  \min_{\{X_\alpha\} \vdash A,\ \{Y_\beta\} \vdash B}\ 
  \sum_{\alpha,\beta}\ 
  \relrk_{\alpha \cup \beta \cup \delta^\pm(\zeta),\, \xi \cup \zeta}(X_\alpha \otimes Y_\beta)
  \\
  &\le
  \max_{\xi,\zeta}\ 
  \min_{\{X_\alpha\} \vdash A,\ \{Y_\beta\} \vdash B}\ 
  \sum_{\alpha,\beta}\ 
  \relrk_{\alpha,\,\delta^\pm(\zeta) \cup \xi}(X_\alpha)\ 
  \relrk_{\beta,\, \delta^\pm(\zeta) \cup \zeta}(Y_\beta)
  \\
  &\le
  \max_{\xi,\zeta}\ 
  \min_{\{X_\alpha\} \vdash A,\ \{Y_\beta\} \vdash B}\ 
  \sum_{\alpha,\beta}\ 
  \relrk_{\alpha,\, \delta^\pm(\zeta) \cup \xi}(X_\alpha)\ 
  n^{-|V_1(G) \,\cap\, V_1(H)|}
  \\
  &= 
  n^{-|V_1(G) \,\cap\, V_1(H)|}\ 
  \max_{\xi,\zeta}\ 
  \min_{\{X_\alpha\} \vdash A}\ 
  \sum_\alpha\ 
  \relrk_{\alpha,\, \delta^\pm(\zeta) \cup \xi}(X_\alpha)
  \\
  &\le
  n^{-|V_1(G) \,\cap\, V_1(H)|
  }\ 
  \max_{\xi,\delta}\ 
  \min_{\{X_\alpha\} \vdash A}\ 
  \sum_\alpha\ 
  \relrk_{\alpha,\, \delta \cup \xi}(X_\alpha)
  \\
  &=
  n^{-|V_1(G) \,\cap\, V_1(H)|
  }\ 
  \rho(A).\qedhere
\]
\end{proof}

\subsection{Set-multilinear formula lower bounds from $\rho$}

The next definition generalizes the flattening operation $\flat$ introduced in Def.\ \ref{df:flat1}.

\begin{df}
For any $G \subseteq \mr{Path}_\Z$ and $A : [n]^{D(G)} \to \F$, let $A^\flat : [n^2]^{E(G)} \to \F$ denote the flattened 
tensor
defined in the natural way: for $x \in [n]^{D(G)}$, we have $A^\flat(y) = A(x)$ where $y \in [n^2]^{E(G)}$ is the element defined by $y_{\{v_{i-1},v_i\}} = \sq{x_{v_{i-1}},x_{v_i}}$ for each $i \in \Z$ such that $\{v_{i-1},v_i\} \in E(G)$.
\end{df}

The analog of Theorem \ref{thm:nc-lower-bound} for (commutative) set-multilinear formulas has  a similar proof based on the inductive definition of $\Lsm$ However, instead of $\log d$ in the exponent of $n$, the relevant parameter becomes $\log \ell(G)$ where $\ell(G)$ is the length of the longest path in $G$.  We record this notation below:

\begin{notation}
For $G \subseteq \Path_\Z$, let $\ell(G)$ denote the length of the longest path in $G$ (i.e.,\ the maximal number of edges in a connected component of $G$).
\end{notation}

\begin{thm}\label{thm:main-lb}
For every $G \subseteq \mr{Path}_\Z$ and $A : [n]^{D(G)} \to \F$,
we have 
\[
  \Lsm(A^\flat) \ge n^{\log \ell(G)}\ \rho(A).
\]
That is, 
any 
set-multilinear arithmetic formula computing the flattened tensor 
$A^\flat : [n^2]^{E(G)} \to \F$ has size at least $n^{1 + \log \ell(G)}\ \rho(A)$.
\end{thm}

\begin{proof}
We argue by induction on $|E(G)|$. The base case $|E(G)| = 1$ is trivial, since $|E(G)|-|V(G)|+1+\log|E(G)| = 0$ and $\rho(A) \le 1$. 

Now assume $|E(G)| \ge 2$ and fix $\{(H_i,K_i,B_i,C_i)\}_i$ with $H_i,K_i \subsetneqq G$ such that $H_i \cup K_i = G$ and $B_i : [n]^{D(H_i)} \to \F$ and $C_i : [n]^{D(K_i)} \to \F$ such that $A = \bigcup_i B_i \otimes C_i$ and $\Lsm(A) = \sum_i\ (\Lsm(B_i) + \Lsm(C_i))$.

Note that for all $i$,
\[
  \ell(G) 
  \le 
  \max\{\ell(H_i),\, \ell(K_i)\} \cdot (|V_1(H_i) \cap V_1(K_i)|+1).
\]
Therefore,
\[
  \log\ell(G)
  \le
  \max\{\log\ell(H_i),\, \log\ell(K_i)\} 
  + |V_1(H_i) \cap V_1(K_i)|.
\]
We now have
\[
  \Lsm(A^\flat) 
  &\ge
  \sum_i\,
  \Big(
     n^{\log\ell(H_i)}\ \rho(B_i) + n^{\log\ell(K_i)}\ \rho(C_i)
  \Big)
  \quad\text{(induction hypothesis)}
  \\
  &\ge
  \sum_i\,
     n^{\max\{\log\ell(H_i),\, \log\ell(K_i)\} \,+\, |V_1(H_i) \,\cap\, V_1(K_i)|}\ \rho(B_i \otimes C_i)
  \quad\text{(Lemma \ref{la:rho-tensor2})}
  \\
  &\ge
  n^{\log \ell(G)}\ 
  \sum_i\
  \rho(B_i \otimes C_i)\\
  &\ge
  n^{\log \ell(G)}\ 
  \rho(\,
  \sum_i\, B_i \otimes C_i)
  \quad\text{(Lemma \ref{la:rho-subadd2})}
  \\
  &=
  n^{\log  \ell(G)}\ \rho(A).\qedhere
\]
\end{proof}

In the special case where $G$ is path of length $d$, Theorem \ref{thm:main-lb} has 
the following corollary, which directly strengthens Theorem \ref{thm:nc-lower-bound} by replacing $\Lnc$ with $\Lsm$.

\begin{cor}
For every $A : [n]^{2d} \to \F$, we have $\Lsm(A^\flat) \ge n^{\log d}\ \rho(A)$. 
\end{cor}

Finally, Theorem \ref{thm:the-main} follows directly from Theorem \ref{thm:nc-lower-bound} via the same observation that 
\[
\PTrank(M) = n^{d+1}\,\rho(\PaddedT{M})
\]
for every $n^d \times n^d$ matrix $M$ and that $\ShiftT{M}=\PaddedT{M}^\flat$.

\section{Upper bounds and candidate hard matrices for PT-rank}\label{sec:VBP}

In this section, we first present some known PT-rank upper bounds for certain families of matrices. These bounds are based on Hrubes' recent subquadratic upper bound for the sum-of-squares problem \cite{hrubes2024subquadratic}. Then, we give some explicit candidate hard matrices that might realize our desired lower bound.

\subsection{Kronecker products of $n\times n$ matrices have low PT-rank}

Recall Hrube\v{s}' upper bound $\SOS(q_n) = O(n^{1.62})$ (Theorem \ref{thm:Hrubes}). 
By Proposition \ref{prop:pt-rank-bil}, this is equivalent to $\PTrank(I_{n^2}) = O(n^{1.62})$. 
This immediately gives the following upper bound on $\PTrank\paren{I_{n^d}}$ for all $d$.

\begin{prop}\label{thm:81}
$\PTrank(I_{n^d})= n^{0.81d + O(1)}$.
\end{prop}
\begin{proof}
We prove the bound $\PTrank\paren{I_{n^d}}= O(n^{0.81d})$ in the case where $d$ is even. (The bound $n^{0.81d+O(1)}$ for odd $d$ follows by monotonicity.)
This bound is given by the following calculation:
    \[
    \PTrank_{[n]^d}(I_{n^d})
    &\le \PTrank_{[n^{d/2}]^2}\paren{I_{(n^{d/2})^2}}
    = O((n^{d/2})^{1.62}) = O(n^{0.81d}).
    \]
The first inequality holds since $\PTrank_{[n]^d}$ can use $\top_\kappa$ for all $\kappa\subseteq [d-1]$, whereas $\PTrank_{[n^{d/2}]^2}$ can only use $\top_\kappa$ for $\kappa\in \{\emptyset,[d/2]\}$. 
\end{proof}

Notice that the PT-rank bound of Proposition \ref{thm:81} involves only two partial transposes, namely $\top_\emptyset$ and $\top_{[d/2]}$. 
Using all $2^d$ partial transposes, we can show a much stronger upper bound:

\begin{prop}\label{thm:07}
$\PTrank(I_{n^d})=n^{O(d^{0.7})}$.
\end{prop}

\begin{proof}
    The construction by Hrube\v{s} shows that 
    \[\prod_{i=1}^d \bigg(\sum_{j=1}^n X_{i,j}^2\bigg) =\prod_{i'=1}^{d/2}\bigg(\sum_{j'=1}^{O\paren{n^{1.62}}} f_{i',j'}^2\bigg),\]
    where $f_{i',j'}$ is a bilinear form in $\{X_{i,(2j'-1)}\}_{i=1}^d$ and $\{X_{i,(2j')}\}_{i=1}^d$. We may repeat the step by considering $f_{i'j'}$ as new variables. Inductively, this gives
    \[\prod_{i=1}^d \bigg(\sum_{j=1}^n X_{i,j}^2\bigg)=\sum_{j'=1}^{s}g_{j'}^2,\]
    where $g_{j'}$ is multilinear in the $d$ sets of variables and \[s=O\paren{n^{\paren{1.62}^{\log_2 d}}}\le n^{O(d^{0.7})}.\]
    The corollary follows from Proposition \ref{prop:pt-rank-bil}. 
\end{proof}

The upper bound above extends to Kronecker products of $n \times n$ matrices, via  $GL(n,\F)^d$-invariance of PT-rank (Lemma \ref{la:invariance}).
Thus, we have proved:

\begin{thm}\label{thm:kron}
If $M$ is a Kronecker product of $d$ $n \times n$ matrices, then $\PTrank(M)$ is at most $n^{O(d^{0.7})}$.
\end{thm}

\subsection{Matrices in $\VBPord$ have low PT-rank}

The Kronecker products of $n\times n$ matrices have low PT-rank. Using this fact, we may derive an upper bound for all matrices whose shifted tensors are in the complexity class $\VBPord$. 

\begin{definition}[Ordered algebraic branching programs]
    Let $f$ be a set-multilinear polynomial over $\F$ of degree $d$ in variable sets $X^{(1)},X^{(2)},\dots, X^{(d)}$. We say that $f$ has an ordered algebraic branching program of width $w$ and size $s$ if
    \[f=v_1^\transpose M_1
     M_2\cdots M_d v_2,\]
    where $v_1\in \F^{w_0}$ and $v_2\in \F^{w_d}$ are vectors whose entries are in $\F$, $M_i\in \F^{w_{i-1}\times w_i}$ are matrices whose entries are linear combinations of variables in $X_i$ for $i\in [d]$, and the following holds: 
    \[w_i\le w\text{ for }0\le i\le [d]\text{ and }\sum_{i=0}^d w_i\le s.\]
\end{definition}

\begin{definition}[$\VBPord$]
    A sequence of set-multilinear polynomials with $n$ variables is in $\VBPord$ if and only if it can be computed by a sequence of ordered algebraic branching programs with size $s=n^{O(1)}$.  
\end{definition}

An ordered algebraic branching program for a tensor $A$ is an ordered algebraic branching program for its associated set-multilinear polynomial $P_A$ (as defined in Section \ref{sec:p_A}). The class $\VBPord$ is defined for tensors similarly. 

Our main result (Theorem \ref{thm:the-main}) shows that sufficiently strong lower bounds on $\PTrank(M)$ 
for explicit matrices $M$ would imply a separation of classes $\VNCone$ and $\mathsf{VNP}$. 
For a more stringent notion of ``explicit'', this reduction could even potentially separate $\VNCone$ from $\mathsf{VP}$ (i.e., arithmetic formulas from circuits). 
In this section, however, we describe a limitation of PT-rank by showing that it cannot separate $\VNCone$ from $\VBPord$ (i.e., arithmetic formulas from ordered branching programs).

\begin{thm}\label{thm:limitation-theorem}
    For any sequence of $n^d\times n^d$ matrices $M$ such that $\mathrm{ShiftedTensor}(M)\in \VBPord$, we have $\PTrank(M)\le n^{0.81d+O(1)}$. 
\end{thm}

\begin{proof}
To simplify notation in the proof, assume $M$ has dimensions $n^{d-1} \times n^{d-1}$  (instead of $n^d \times n^d$), where $d$ is odd.
Then, $\ShiftT{M}$ is a $[n^2]^d \to \F$ tensor.

Let $A$ be the unique $[n]^{2d} \to \F$ tensor with $A^\flat = \ShiftT{M}$. That is, for all $a_1,\dots,a_{2d} \in [n]$, we have
\[
  A(a_1,a_2,\dots,a_{2d-1},a_{2d})
  =
  \ShiftT{M}(\sq{a_1,a_2},\dots,\sq{a_{2d-1},a_{2d}})
  =
  M_{(a_3,a_5,\dots, a_{2d-1}),(a_2, a_4,\dots ,a_{2d-2})}.
\]

Let $f=v_1^\transpose M_1M_2\cdots M_dv_2$ be the ordered branching program for $\ShiftT{M}$. Then, $P_{\ShiftT{M}}$ can be expressed as \[
P_{\ShiftT{M}}=\sum_{(i,j)\in [w]^2}\paren{v_1^\transpose M_1M_2\cdots M_{\floor{d/2}}}(i)\cdot M_{\ceil{d/2}}(i,j)\cdot \paren{M_{\ceil{d/2}+1}\cdots M_d v_2}(j).
\]
By definition, each $M_{\ceil{d/2}}(i,j)$ is a linear combination of variables in $X^{(\ceil{d/2})}$, which contains $n^2$ variables. Let $\coeff_{\chevron{a,b}}\paren{M_{\ceil{d/2}}(i,j)}$ be the coefficient of $X_{\chevron{a,b}}^{(\ceil{d/2})}$ in $M_{\ceil{d/2}}(i,j)$. Then, 
\begin{multline*}
P_{\ShiftT{M}}=\sum_{(i,j)\in [w]^2, \chevron{a,b}\in [n^2]}\paren{v_1^\transpose M_1M_2\cdots M_{\floor{d/2}}}(i)\\ \cdot \coeff_{\chevron{a,b}}\paren{M_{\ceil{d/2}}(i,j)} \cdot X_{\chevron{a,b}}^{(\ceil{d/2})}\cdot \paren{M_{\ceil{d/2}+1}\cdots M_d v_2}(j).
\end{multline*}
Note that each $\paren{v_1^\transpose M_1M_2\cdots M_{\floor{d/2}}}(i)$ is a set-multilinear polynomial in $X^{(1)},\dots, X^{(\floor{d/2})}$, and $\paren{M_{\ceil{d/2}+1}\cdots M_d v_2}(j)$ is a set-multilinear polynomial in $X^{(\ceil{d/2}+1)},\dots, X^{(d)}$. Therefore, the expression above induces a decomposition of $\Mat_{\{1,\dots,d \}, \{d+1,\dots, 2d \}}\paren{A}$ into $w^2\cdot n^2=n^{O(1)}$ rank $1$ matrices. 
As a result, \[\rank\paren{\Mat_{\{1,\dots,d \}, \{d+1,\dots, 2d \}}\paren{A}}\le n^{O(1)}.\]
Recall that when creating the shifted tensors, indices $1$ and $2d$ are padded. We may therefore consider the submatrix ignoring these indices $\Mat_{\{2,3,\dots,d\}, \{d+1,\dots, 2d-1 \}}\paren{A}$. Formally, define the matrix by
\begin{multline*}
    \paren{\Mat_{\{2,3,\dots,d\}, \{d+1,\dots, 2d-1 \}}\paren{A}}_{(a_2,\dots, a_d),(a_{d+1},\dots, a_{2d-1})}\\
    =\paren{\Mat_{\{1,2,\dots,d\}, \{d+1,\dots, 2d-1, 2d\}}\paren{A}}_{(1,a_2,\dots, a_d),(a_{d+1},\dots, a_{2d-1}, 1)}
\end{multline*}
We then have
\[\rank\paren{\Mat_{\{2,\dots,d\}, \{d+1,\dots, 2d-1 \}}\paren{A}}\le n^{O(1)}.\]
Therefore, there are tensors $B_1,C_1,\dots,B_w,C_w : [n]^{d-1} \to \F$ such that
\[\Mat_{\{2,\dots,d \}, \{d+1,\dots, 2d-1 \}}\paren{A}=\sum_{l=1}^w \mr{vec}(B_l) \mr{vec}(C_l)^\transpose\]
where $\mr{vec}(B_l),\mr{vec}(C_l) \in \F^{n^d}$ are the vectorizations of $B_l$ and $C_l$, namely flattening $B_l$ and $C_l$ in a way that is consistent to $A$. We now have
\begin{multline*}
M=\Mat_{\{3,5,\dots, 2d-1\},\{2,4,\dots, 2d-2\}}(A)\\=\sum_{l=1}^w \Mat_{\{3,5,\dots, d\},\{2,4,\dots, d-1\}}(B_l)\kron \Mat_{\{d+2,d+4,\dots, 2d-1\},\{d+1,d+3,\dots, 2d-2\}}(C_l).
\end{multline*}
Using subadditivity of PT-rank and tensor product properties from Lemma \ref{la:invariance}, we may upper bound the PT-rank of the flattening of $A$ as follows: 
\begin{align*}
    &\PTrank_{[n]^{d-1}}\paren{M}\\&=\PTrank_{[n]^{d-1}} \Big(\sum_{l=1}^w \Mat_{\{3,5,\dots, d\},\{2,4,\dots, d-1\}}(B_l)\kron \Mat_{\{d+2,d+4,\dots, 2d-1\},\{d+1,d+3,\dots, 2d-2\}}(C_l)\Big)\\
    &\le \PTrank_{[n^{(d-1)/2}]^2} \Big(\sum_{l=1}^w \Mat_{\{3,5,\dots, d\},\{2,4,\dots, d-1\}}(B_l)\kron \Mat_{\{d+2,d+4,\dots, 2d-1\},\{d+1,d+3,\dots, 2d-2\}}(C_l)\Big)\\
    &\le \sum_{l=1}^w \PTrank_{[n^{(d-1)/2}]^2} \Big(\Mat_{\{3,5,\dots, d\},\{2,4,\dots, d-1\}}(B_l)\kron \Mat_{\{d+2,d+4,\dots, 2d-1\},\{d+1,d+3,\dots, 2d-2\}}(C_l)\Big)\\
    &\le \sum_{l=1}^w \PTrank_{[n^{(d-1)/2}]^2} (I_{n^{(d-1)/2}}\kron I_{n^{(d-1)/2}})\\
    &\le n^{O(1)}\cdot n^{0.81d+O(1)}
    =n^{0.81d+O(1)}.\qedhere
\end{align*}
\end{proof}

\subsection{Candidate hard matrices}

We now propose a natural family of $n^d \times n^d$ matrices over $\F=\C$ which are plausible candidates for having nearly maximal PT-rank. 

\begin{definition}
    Let $d$ be an even integer. Let $n$ be a prime power such that $n>2d$. Let $T$ be an $d \times d$ matrix over the prime field $\F_n$. Let $\omega=e^{2\pi i/n}$ be a primitive $n$th root of unity. Then, define the $n^d\times n^d$ matrix $W_T$ over $\C$ by
    \[(W_T)_{(i_1,\dots, i_d),(j_1,\dots, j_d)}=\omega^{(i_1,\dots, i_d)T(j_1,\dots, j_d)^\top}.\]
\end{definition}

With the definition above, we first consider three choices $T_1,T_2,T_3$ that don't work, that is, where $W_T$ has small PT-rank. 

\paragraph{Identity matrix. } For $T_1=I_d$, we have \[(W_{T_1})_{(i_1,\dots, i_d),(j_1,\dots, j_d)}=\omega^{(i_1,\dots, i_d)(j_1,\dots, j_d)^\top}.\]
We may define $n\times n$ matrices $W_{T_1,l}$ for $l\in [d]$ by 
$(W_{T_1, l})_{i_l,j_l}=\omega^{i_lj_l}$.
Then, we have
$W_{T_1}=W_{T_1, 1}\kron
\cdots \kron
W_{T_1, d}$.
By Theorem \ref{thm:kron}, this implies 
$
\PTrank(W_{T_1})=n^{O(d^{0.7})}.$

\paragraph{Cyclic matrix. } Consider the cyclic permutation matrix $T_2$ defined by \[(T_2)_{a,b}=1[a+1 \equiv b\mod{d}].\]
For this choice of $T_2$, we have \[(W_{T_2})_{(i_1,\dots, i_d),(j_1,\dots, j_d)}=\omega^{(i_1,\dots, i_{d-1}, i_d)(j_2,\dots, j_d, j_1)^\top}.\]
Let $\kappa=\{1,3,5,\dots, d-1\}$. Then, by definition of partial transpose, we have
\[(W_{T_2}^{\top_\kappa})_{(j_1,i_2,\dots, j_{d-1}, i_d), (i_1,j_2,\dots, i_{d-1}, j_d)}=(W_{T_2})_{(i_1,\dots, i_d),(j_1,\dots, j_d)}=\omega^{(i_1,\dots, i_{d-1}, i_d)(j_2,\dots, j_d, j_1)^\top}.\]
Define $n^d$ vectors $u$ and $v$ by
\[
u_{(j_1,i_2,\dots, j_{d-1})}=\omega^{(i_2, i_4,\dots, i_d)(j_1,j_3,\dots, j_{d-1})^\top}, 
v_{(i_1,j_2,\dots, i_{d-1})}=\omega^{(i_1, i_3,\dots, i_{d-1})(j_2,j_4,\dots, j_d)^\top}.
\]
We then have $W_{T_2}^{\top_\kappa}=uv^\top$, which means it has rank $1$. Therefore, $\PTrank(W_{T_2})=1$. 

\paragraph{Upper triangular matrix. } Consider the upper triangular matrix $T_3$ defined by 
\[(T_3)_{a,b}=1[a\le b].\]
For this choice of $T_3$, we have
$
(W_{T_3})_{(i_1,\dots, i_d),(j_1,\dots, j_d)}=\omega^{\sum_{k\le l}i_kj_l}.$

Let $A$ be the unique $[n]^{2d}$ tensor such that $A^\flat=\ShiftT{W_{T_3}}$. Then, 
\[A(p,i_1,j_1,i_2, \dots, j_d,q)=(W_{T_3})_{(i_1,\dots, i_d),(j_1,\dots, j_d)}.\]
Then, it can be observed that $\rank(\Mat_{\{1,\dots, d+1\},\{d+2,\dots, 2d+2\}})\le n^2$. (In fact, $A^\flat\in \VBPord$.) Therefore, $\PTrank(W_{T_3})\le n^{0.81d+O(1)}$ by Theorem \ref{thm:limitation-theorem}.\bigskip

The three matrices $T_1,T_2,T_2$ considered above give rise to $W_T$ with small PT-rank. 
We now give some evidence that a {\em Cauchy matrix} (in which every submatrix has full rank) might be a good choice of $T$.

\begin{proposition}\label{prop:fullrank}
Let $T$ be a Cauchy matrix. For every $\kappa\subseteq[d]$, the matrix $W_T^{\top_\kappa}$ has full rank $n^d$.
\end{proposition}

\begin{proof}
Fix $\kappa$.
Write $\kappa^c=[d]\setminus \kappa$ and block-partition
\[
T=\begin{pmatrix}T_{\kappa^c,\kappa^c}&T_{\kappa^c,\kappa}\\ T_{\kappa,\kappa^c}&T_{\kappa,\kappa}\end{pmatrix}.
\]
Let rows of $(W_T^{\top_\kappa})$ be indexed by $(\alpha,\beta)=(i_{\kappa^c},\,j_\kappa)$ and columns by $(\gamma,\delta)=(i_\kappa,\,j_{\kappa^c})$.
Then
\begin{align*}
(W_T^{\top_\kappa})_{(\alpha,\beta),(\gamma,\delta)}
&= \omega^{\,\alpha T_{\kappa^c,\kappa^c}\delta^\top + \alpha T_{\kappa^c,\kappa}\beta^\top + \gamma T_{\kappa,\kappa^c}\delta^\top + \gamma T_{\kappa,\kappa}\beta^\top}\\
&=\ \underbrace{\omega^{\,\alpha T_{\kappa^c,\kappa^c}\delta^\top}}_{\text{DFT block}}\cdot
   \underbrace{\omega^{\,\gamma T_{\kappa,\kappa}\beta^\top}}_{\text{DFT block}}\cdot
   \underbrace{\omega^{\,\alpha T_{\kappa^c,\kappa}\beta^\top + \gamma T_{\kappa,\kappa^c}\delta^\top}}_{\text{permutation of indices}}.
\end{align*}
Note that the last term is just repeated scaling of individual rows and columns, which doesn't change the rank. Since $T_{\kappa^c,\kappa^c}$ and $T_{\kappa,\kappa}$ are invertible (Cauchy), the first two terms are the Kronecker product of DFT matrices of sizes $n^{|\kappa^c|}$ and $n^{|\kappa|}$, which are both nonsingular. Therefore, $W_T^{\top_\kappa}$ is full-rank. 
\end{proof}

\begin{definition}
    For $\lambda \subseteq [d]$, let $W_T^{[\lambda]}$ be the $n^{2|\lambda|} \times n^{2(d-|\lambda|)}$ matrix defined by
\[
  (W_T^{[\lambda]})_{(i_\lambda,j_\lambda), (i_{\lambda^c},j_{\lambda^c})} \;=\; 
  (W_T)_{(i_1,\dots,i_d), (j_1,\dots,j_d)}.
\]
\end{definition}

\begin{proposition}
Let $T$ be a Cauchy matrix. For every $\lambda\subseteq[d]$, the matrix $W_T^{[\lambda]}\in\mathbb{C}^{\,n^{2|\lambda|}\times n^{2(d-|\lambda|)}}$ has full rank:
\[
\rank(W_T^{[\lambda]})
\;=\; n^{\,2\min\{|\lambda|,\,d-|\lambda|\}}.
\]
\end{proposition}

\begin{proof}
Write $\lambda^c=[d]\setminus \lambda$ and block-partition
\[
T=\begin{pmatrix}T_{\lambda^c,\lambda^c}&T_{\lambda^c,\lambda}\\ T_{\lambda,\lambda^c}&T_{\lambda,\lambda}\end{pmatrix}.
\]

Let the matrix $W_T^{[\lambda]}$ be indexed by $(u,v)=(i_\lambda, j_\lambda), (x,y)=(i_{\lambda^c}, j_{\lambda^c})$. Then
\begin{align*}
(W_T^{[\lambda]})_{(u,v),(x,y)}
&= \omega^{\,x T_{\lambda^c,\lambda^c}y^\top + x T_{\lambda^c,\lambda}v^\top + u T_{\lambda,\lambda^c}y^\top + u T_{\lambda,\lambda}v^\top}\\
&=\omega^{\,x T_{\lambda^c,\lambda}v^\top}\cdot
   \omega^{\,u T_{\lambda,\lambda^c}y^\top}\cdot
   \underbrace{\omega^{\,x T_{\lambda^c,\lambda^c}y^\top + u T_{\lambda,\lambda}v^\top}}_{\text{permutation of indices}}.
\end{align*}
Again, note that the last term is just repeated scaling of individual rows and columns. Therefore, it suffices to consider the first two terms. The matrix $(W_T)^{[\lambda]}$ is just the Kronecker product of the matrices determined by these terms. By symmetry, it suffices to bound the rank of one of them. 

Assume without loss of generality that $\abs{\lambda}\leq \abs{\lambda^c}$. Then, we consider the matrix $[\omega^{\,u T_{\lambda,\lambda^c}y^\top}]$ indexed by $(u,y)$. Then, let $c_u$ be the row indexed by $u$. Consider $u\neq u'$. We then have
\[\langle c_u, c_{u'}\rangle=\sum_{y} \omega^{(u-u')T_{\lambda,\lambda^c} y^\top}=0\]
since $T_{\lambda,\lambda^c}$ is full rank. Therefore, both $[\omega^{\,u T_{\lambda,\lambda^c}y^\top}]$ and $[\omega^{\,x T_{\lambda^c,\lambda}v^\top}]$ have full-rank $n^{\,\min\{|\lambda|,\,d-|\lambda|\}}$.
\end{proof}

The two propositions above disable the only two upper bound techniques we know: taking a single partial transpose, and finding some flattening of $K_T$ with small rank in order to apply Hrube\v{s}' construction.

\bibliographystyle{abbrv}


\bibliography{PT-rank-refs.bib}

\end{document}